\documentclass{aa}
\usepackage{graphicx}
\usepackage{subcaption}
\usepackage[colorlinks=true, citecolor=blue, linkcolor=blue, urlcolor=black]{hyperref}
\usepackage[varg]{txfonts}
\usepackage{appendix}
\usepackage{comment}
\usepackage{mathrsfs}
\usepackage{amsmath}
\usepackage{multicol}
\usepackage{multirow,tabularx}
\usepackage{longtable}
\DeclareUnicodeCharacter{2212}{-}
\newcommand{\RNum}[1]{\uppercase\expandafter{\romannumeral #1\relax}}

\begin{document}

   \title{Discovery of 24 radio-bright quasars at $4.9 \leq z \leq6.6$ using low-frequency radio observations \thanks{Based on observations obtained with the Hobby-Eberly Telescope (HET), which is a joint project of the University of Texas at Austin, Pennsylvania State University, Ludwig-Maximillians-Universitaet Muenchen, and Georg-August Universitaet Goettingen. The HET is named in honor of its principal benefactors, William P. Hobby and Robert E. Eberly.}}
   \author{A. J. Gloudemans \inst{\ref{inst1}}
   \and K. J. Duncan \inst{\ref{inst_edin}} 
   \and A. Saxena \inst{\ref{inst_ucl}}
   \and Y. Harikane \inst{\ref{inst_tokyo}}
   \and G. J. Hill \inst{\ref{inst_mcdonald},\ref{inst_texas}}
   \and G. R. Zeimann \inst{\ref{inst_het}}
   \and H. J. A. R\"{o}ttgering \inst{\ref{inst1}}
   \and D. Yang \inst{\ref{inst1}}
   \and P. N. Best \inst{\ref{inst_edin}}
   \and E. Ba\~{n}ados \inst{\ref{inst_heidel}}
   \and A. Drabent \inst{\ref{inst_thur}}
   \and M. J. Hardcastle \inst{\ref{inst_hert}}
   \and J. F. Hennawi \inst{\ref{inst1},\ref{inst_cal}}
   \and G. Lansbury \inst{\ref{inst_eso}}
   \and M. Magliocchetti \inst{\ref{inst_inaf}}
   \and G. K. Miley \inst{\ref{inst1}}
   \and R. Nanni \inst{\ref{inst1}}
   \and T. W. Shimwell \inst{\ref{inst1}, \ref{inst_astron}}
   \and D. J. B. Smith \inst{\ref{inst_hert}}
   \and B. P. Venemans \inst{\ref{inst1}}
   \and J. D. Wagenveld \inst{\ref{inst_bonn}}
   }
   
   \institute{Leiden Observatory, Leiden University, PO Box 9513, 2300 RA Leiden, The Netherlands \\ e-mail: gloudemans@strw.leidenuniv.nl\label{inst1} 
   \and Institute for Astronomy, Royal Observatory, Blackford Hill, Edinburgh, EH9 3HJ, UK\label{inst_edin} \and Department of Physics and Astronomy, University College London, Gower Street, London WC1E 6BT, UK \label{inst_ucl} \and Institute for Cosmic Ray Research, the University of Tokyo, 5-1-5 Kashiwa-no-Ha, Kashiwa City, Chiba, 277-8582, Japan \label{inst_tokyo} \and McDonald Observatory, University of Texas at Austin, 2515 Speedway, Stop C1402, Austin, TX 78712, USA \label{inst_mcdonald} \and Department of Astronomy, University of Texas at Austin, 2515 Speedway, Stop C1400, Austin, TX 78712, USA \label{inst_texas} \and Hobby Eberly Telescope, University of Texas, Austin, TX 78712, USA\label{inst_het} \and Max Planck Institut f\"ur Astronomie, Königstuhl 17, D-69117, Heidelberg, Germany\label{inst_heidel} \and Th\"uringer Landessternwarte, Sternwarte 5, D-07778 Tautenburg, Germany\label{inst_thur} \and Centre for Astrophysics Research, Department of Physics, Astronomy and Mathematics, University of Hertfordshire, College Lane, Hatfield AL10 9AB, UK\label{inst_hert} \and Department of Physics, University of California, Santa Barbara, CA 93106-9530, USA\label{inst_cal} \and European Southern Observatory, Karl-Schwarzschild str. 2, D-85748 Garching bei München, Germany\label{inst_eso} \and INAF-IAPS, Via del Fosso del Cavaliere 100, 00133 Rome, Italy\label{inst_inaf} \and ASTRON, Netherlands Institute for Radio Astronomy, Oude Hoogeveensedijk 4, 7991 PD, Dwingeloo, The Netherlands\label{inst_astron}  \and Max-Planck Institut für Radioastronomie, Auf dem Hügel 69, 53121 Bonn, Germany\label{inst_bonn}}

   \date{Received: 18 August 2022 / Accepted: 28 September 2022}
 
 \abstract{High redshift quasars ($z>5$) that also shine brightly at radio wavelengths are unique signposts of supermassive black hole activity in the early universe. However, bright radio sources at $z \ge 5$ are extremely rare and therefore we have started a campaign to search for new high-$z$ quasars by combining an optical dropout selection driven by the $g$, $r$, and $z$ bands from the Dark Energy Spectroscopic Instrument (DESI) Legacy Imaging Surveys with low-frequency radio observations from the LOFAR Two-metre Sky Survey (LoTSS). Currently, LoTSS covers a large fraction of the northern sky ($\sim$5720 deg$^2$) to such a depth (median noise level $\sim83$ $\mu$Jy beam$^{-1}$) that about 30\% of the general quasar population is detected $-$ which is a factor of 5-10 more than previous large sky radio surveys such as NVSS and FIRST, respectively. In this paper, we present the discovery of 20 new quasars (and the independent confirmation of four) between $4.9 \leq z \leq6.6$. Out of the 24 quasars, 21 satisfy the traditional radio-loudness criterion of $R=f_{5\text{GHz}}/f_{4400 \text{\AA}} > 10$, with the full sample spanning $R\sim$6-1000, thereby more than doubling the sample of known radio-loud quasars at $z \ge 5$. Our radio detection requirement strongly decreases the contamination of stellar sources and allows one to select these quasars in a broad redshift range. Despite selecting our quasar candidates using fewer and less conservative colour restrictions, both the optical and near-infrared colours, Ly$\alpha$ emission line properties, and dust reddening, $E(B-V)$, measurements of our quasar sample do not deviate from the known radio-quiet quasar population, suggesting similar optical quasar properties of the radio-loud and radio-quiet quasar population at high-$z$. Our campaign demonstrates the potential for discovering new high-$z$ quasar populations through next generation radio continuum surveys.}


\maketitle

\section{Introduction}
\label{sec:introduction}

High redshift quasars ($z>5$) are powerful tools for studying supermassive black hole (SMBH) formation and evolution and the phase transition from neutral to ionised hydrogen during the Epoch of Reionization (EoR). In the past few decades, hundreds of quasars at $z>5$ have been discovered in various optical and infrared surveys (e.g. \citealt{Willott2007,Venemans2013ApJ...779...24V,Banados2016ApJS..227...11B,Matsuoka2016ApJ...828...26M,WangF2017,Miyazaki2018,dey2019AJ....157..168D}), which have been used to constrain SMBH growth (e.g. \citealt{Onoue2019ApJ...880...77O}) and trace neutral hydrogen in the EoR by measuring, for example, the Ly$\alpha$ damping wing (e.g. \citealt{Mortlock2011Natur.474..616M, schroeder2013evidence, Greig2017MNRAS.466.4239G, Banados2018Natur.553..473B}). 

Within this population, the subset of quasars that are also radio bright have been suggested to host more massive SMBHs at lower redshift \citep{McLure2004MNRAS.353L..45M} and are important laboratories for studying the dominant physical mechanisms that contributed to the formation and subsequent quenching of massive galaxies, for example, gas accretion, galaxy merging, SMBH growth, and active galactic nucleus (AGN) feedback \citep{Miley2008A&ARv..15...67M, Saxena2017MNRAS.469.4083S,Shao2020A&A...641A..85S, Yamashita2020AJ....160...60Y,Ichikawa2021ApJ...921...51I, Uchiyama2022ApJ...926...76U}. Observations of the radio jets launched by these quasars can provide constraints on the timescales of early AGN activity by studying their synchrotron emission \citep{Saxena2017MNRAS.469.4083S, Shao2020A&A...641A..85S}. In addition, the physical mechanisms by which quasars are triggered and evolve over the Universe's lifetime can be studied through comparisons of radio properties, such as radio sizes and spectral shapes, at high and low redshifts \citep{Hook1998ASPC..146...17H, Banados2015ApJ...804..118B, Gloudemans2021A&A...656A.137G}.

The most radio-bright high-$z$ quasars at low frequencies ($\gtrsim10$ mJy) can, in principle, be used to directly measure the neutral hydrogen content in the EoR by measuring the 21-cm absorption feature, because this line remains optically thin at $z>6$ \citep{carilli2002ApJ...577...22C, Mack2012MNRAS.425.2988M, ciardi2015aska.confE...6C} and is shifted below $200$ MHz at $z\gtrsim6$. Such a measurement will have profound consequences for cosmology, since the rate and patchiness of reionization and the contributors are still heavily debated (see e.g. \citealt{Pentericci2014ApJ...793..113P, Sobral2016hst..prop14699S, Naidu2020ApJ...892..109N}). A recent study of \cite{Bosman2022MNRAS.514...55B}, using deep optical/NIR spectra of 30 high-$z$ quasars, has shown that residual neutral hydrogen in the intergalactic medium (IGM) persists until $z\sim5.3$, which signifies a later end of cosmic reionization than previously found (e.g. \citealt{Becker2001AJ....122.2850B,Fan2001AJ....122.2833F, Fan2006ARA&A..44..415F,McGreer2013}). To study the cosmic reionization process universally via the 21-cm line, multiple sight lines are necessary, but progress has been halted by the small number of known bright high-$z$ radio sources, with only a select number of radio-loud sources lying at $z>6$ \citep{McGreer2006ApJ...652..157M, Willott2010AJ....139..906W, Belladitta2020A&A...635L...7B, Banados2021ApJ...909...80B, Endsley2022arXiv220600018E}. 

The new generation of low-frequency radio telescopes, such as the Low Frequency Array (LOFAR; \citealt{vanHaarlem2013A&A...556A...2V}), are able to cover large areas of the sky with sensitivities in the $\mu$Jy regime, which has enabled high-$z$ quasars to be systematically studied at low frequencies for the first time. The recently released LOFAR Two-Metre Sky Survey second data release (LoTSS-DR2; \citealt{Shimwell2022A&A...659A...1S}) is a low-frequency radio survey of the northern sky at 120-168 MHz with a median root mean square (RMS) sensitivity of $\sim83\ \mu$Jy beam$^{-1}$ and a resolution of 6$\arcsec$ covering over 5720 deg$^2$. With this sensitivity, the survey is reaching more than an order of magnitude deeper (with $\sim 4\times$ better angular resolution) than other large sky surveys at 150 MHz such as the TIFR GMRT Sky Survey (TGSS; \citealt{Intema2017A&A...598A..78I}). \cite{Gloudemans2021A&A...656A.137G} found that 36\% of the known high-$z$ quasars are tentatively detected at $>2\sigma$ significance in LoTSS-DR2, showing the potential of using the new generation radio surveys to study this population. 

In this work, we explore the potential of using LOFAR to discover new radio-loud (RL) quasars above $z>5$. In principle, radio detections drastically reduce the number of stellar contaminants often found in optical and near-infrared (NIR) high-$z$ quasar searches (e.g. \citealt{Hewett2006MNRAS.367..454H, Findlay2012MNRAS.419.3354F, Banados2016ApJS..227...11B}) and therefore allow the exploration of new parameter spaces that were previously inaccessible due to the high contamination by stellar objects (see e.g. \citealt{Banados2015ApJ...804..118B}). Typically, high-$z$ quasar candidates are selected in optical and NIR photometric surveys by probing the Lyman break and redward continuum. In this work we take a similar approach, but select only candidates with LOFAR detections, which allow us to search for quasars in a broad redshift range without requiring the full set of optical dropout bands and without drastically increasing the number of contaminants. Recently, \cite{Wagenveld2022A&A...660A..22W} demonstrated the potential of finding quasars outside the classical optical colour space by discovering a quasar at $z=5.7$ with an optical colour of $i-z = 1.4$, which is below the standard colour selection of $i-z \geq$ 1.5 or 2.0 (e.g. \citealt{Banados2016ApJS..227...11B, Matsuoka2016ApJ...828...26M}), using a probabilistic candidate selection technique. This probabilistic candidate selection has also been performed by e.g. \cite{Mortlock2012MNRAS.419..390M}, \cite{ Matsuoka2016ApJ...828...26M}, and \cite{Nanni2022MNRAS.515.3224N} and has been proven to be efficient in identifying high-$z$ quasars without using traditional discrete colour cuts. 

Recent wide-area high-$z$ quasar searches have utilised the Sloan Digital Sky Survey (SDSS; \citealt{York2000AJ....120.1579Y}) and PANSTARRS-1 (PS1; \citealt{Chambers2016arXiv161205560C}) to select high-$z$ quasars. However, the Dark Energy Spectroscopic Instrument (DESI) Legacy Imaging Surveys (\citealt{dey2019AJ....157..168D}; hereafter referred to as `Legacy Surveys') are reaching at least 1 magnitude deeper in the optical $g-,\ r-,$ and $z$-band than PS1. By requiring a radio detection, dropout candidates between the $r$- and $z$-band can be identified without the need for an $i$-band detection to decrease the fraction of contaminants, therefore allowing us to search for high-$z$ quasars in this deep large sky survey. A recent study of \cite{Wang2019ApJ...884...30W} has also utilised the Legacy Surveys in combination with other optical and NIR surveys to search for new high-$z$ quasars.

In addition to offering potential new probes of the process of reionization, improved samples of the most RL quasars at $z>5$ also enable statistical studies of the radio-loudness evolution, influence of the radio jet on the host galaxy, and quasar environment at high redshift. In this paper, we aim to constrain the relation between the radio luminosity and rest-frame UV properties such as Ly$\alpha$ emission. Previous studies of high-$z$ radio galaxies (HzRGs) have found either a strong positive correlation between the Ly$\alpha$ and radio luminosity \citep{Jarvis2001MNRAS.326.1563J} or a weak correlation, consistent with no correlation \citep{saxena2019MNRAS.489.5053S}. However, these are based on biased samples of luminous radio galaxies with ultra-steep radio spectral indices and strong Ly$\alpha$ emission. Furthermore, we aim to investigate potential differences in the rest-frame UV properties of the RL and radio-quiet (RQ) high-$z$ quasar population, since the existence of a radio-loudness dichotomy is still heavily debated (see e.g. \citealt{Ivezic2002AJ....124.2364I, Cirasuolo2003MNRAS.346..447C, White2007ApJ...654...99W, Zamfir2008MNRAS.387..856Z, Balokovic2012ApJ...759...30B, Beaklini2020MNRAS.497.1463B,  gurkan2019A&A...622A..11G, Macfarlane2021}). Finally, the dust reddened red quasar population has been shown to have an enhanced fraction of radio-loud quasars (e.g. \citealt{Fawcett2020MNRAS.494.4802F, Rosario2020MNRAS.494.3061R, Glikman2022ApJ...934..119G}). Therefore, in this study we aim to investigate whether our radio selection combined with loose optical colour criteria results in an increased fraction of red quasars.

This paper is structured as follows. In Section~\ref{sec:candidate_selection}, we describe the high-$z$ quasar candidate selection and the optical and radio observations on which this is based. In Section~\ref{sec:spectroscopic_obs} we provide details on the conducted spectroscopic observations of our candidates, and in Section~\ref{sec:new_quasars} we show their spectra (we discuss the non-quasar contaminants that our selection procedure produces in Appendix \ref{sec:appendix_contaminants}). We present the optical and low-frequency radio properties of the newly discovered high-$z$ quasar sample in Section~\ref{sec:quasar_sample}. Finally, in Section~\ref{sec:conclusions} we summarise the results together with future prospects. Throughout this work, we assume a $\Lambda$-CDM cosmology with H$_{0}$= 70 km s$^{-1}$ Mpc$^{-1}$, $\Omega_{M}$ = 0.3, and $\Omega_{\Lambda}$ = 0.7 and use the AB magnitude system \citep{Oke1983ApJ...266..713O}.

\section{Candidate selection}
\label{sec:candidate_selection}

Our high-$z$ quasar candidates have been selected by a traditional dropout selection technique using the Legacy Surveys data combined with radio detections from LoTSS-DR2. In addition, we exploit the photometric redshifts (photo-$z$s) generated by \cite{Duncan2022MNRAS.512.3662D} and apply our own spectral energy distribution (SED) fitting to ensure a reliable high-$z$ candidate sample. Here, we describe the specifics of the surveys and selection criteria used in this work to create our quasar candidate sample.

\subsection{DESI Legacy Imaging Surveys}
\label{subsec:legacy_surveys}

The DESI Legacy Imaging Survey DR8 is an optical imaging survey providing \textit{g}, \textit{r}, and \textit{z}-band imaging covering over 19,000 deg$^2$ of the sky. This survey is the deepest wide-area optical survey to date and reaches median $5\sigma$ depths of 24.6, 24.0, and 23.3 magnitude in the \textit{g}, \textit{r}, and \textit{z}-band filter, respectively. In addition to the optical photometry, the Legacy Surveys catalogues include mid-infrared photometry from the Wide-field Infrared Survey Explorer (WISE; \citealt{wright2010AJ....140.1868W}) at wavelengths of 3.4, 4.6, 12, and 22 $\mu$m from the W1, W2, W3, and W4 bands, respectively. We refer to \cite{dey2019AJ....157..168D} for details on the photometric calibrations and measurements. The combination of depth and wide sky coverage of the Legacy Surveys together with the WISE photometry is well suited for discovering quasars at $z>5$.

\subsubsection{Photometric redshifts}
To create our initial sample, we select optical sources from the Legacy Surveys classified as high-$z$ galaxy candidates by the photometric redshift estimates of \cite{Duncan2022MNRAS.512.3662D}. These photo-$z$s have been determined with a machine-learning approach, using sparse Gaussian processes augmented with Gaussian mixture models, together with cost-sensitive learning weight calculations. By combining these two methods, the photo-$z$ estimates have been shown to be reliable for both the general galaxy population and rare or extreme populations at $4<z<7$. Adding spectroscopically confirmed quasars to the training sample and training on them separately (with a measured outlier fraction of OLF$_{0.15}=2.47\%$) makes this approach highly useful for high-$z$ quasar searches (see \citealt{Duncan2022MNRAS.512.3662D}).

From the catalogues created by \cite{Duncan2022MNRAS.512.3662D}, we selected optical sources with photo-$z$ estimates of $z_{\text{phot}} > 3.5$ within the LoTSS-DR2 footprint. To further ensure reliable photometry without blending of flux from neighbouring sources, the \texttt{fracflux} in the \textit{r} and \textit{z}-band filter, which is defined by the profile-weighted fraction of the flux from other sources divided by the total flux, was set to be smaller than \texttt{$0.2$}. In addition, we set \texttt{maskbits=0} to ensure sources are not overlapping with the Legacy masked regions.

\subsubsection{Colour selection criteria}
\label{subsec:colour_selection}
\begin{figure}
\centering
   \includegraphics[width=\columnwidth, trim={0.0cm 0cm 0cm 0.0cm}, clip]{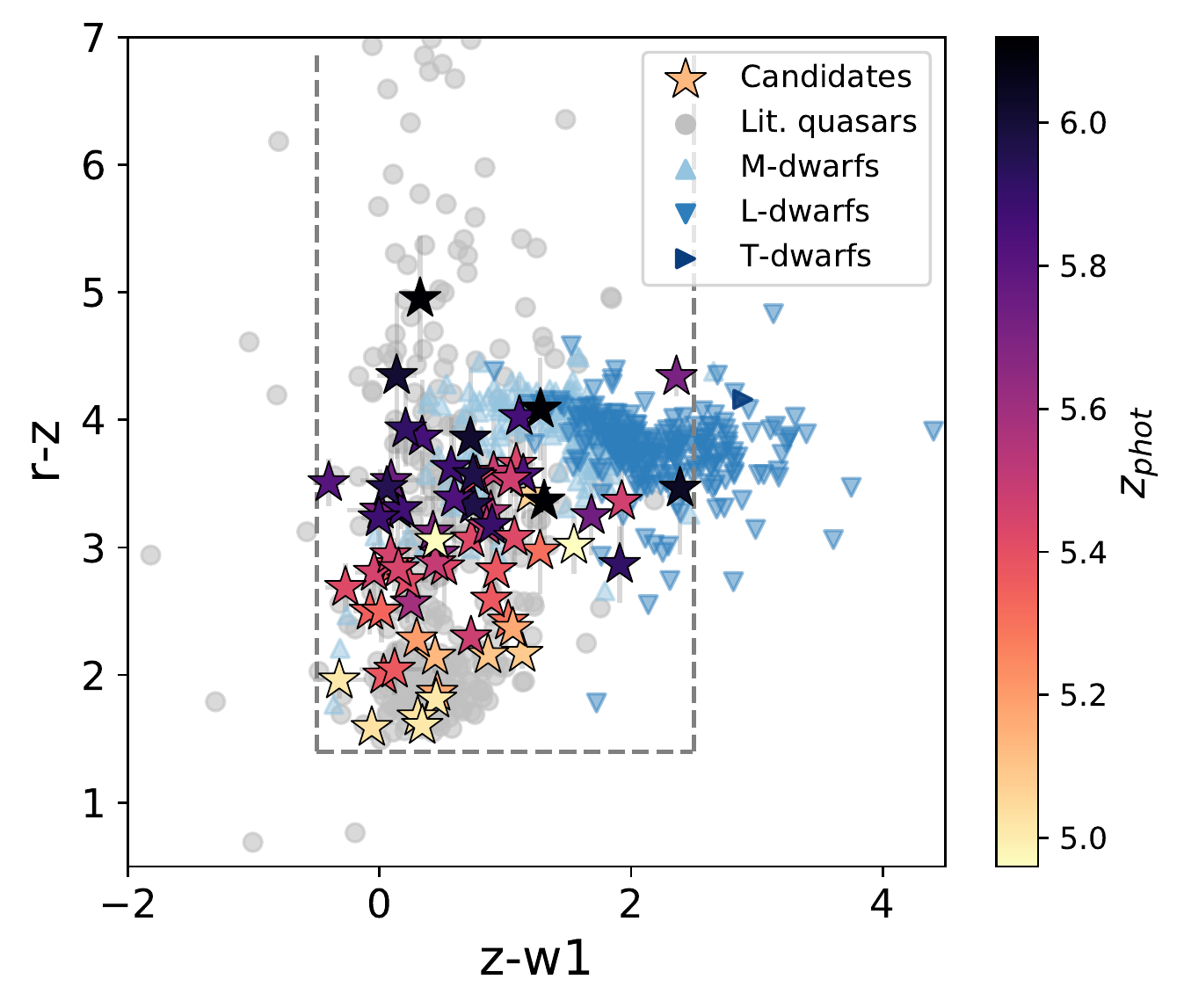}
     \caption{Optical and NIR colour selection criteria (given by the dashed lines) of our spectroscopically observed quasar candidates compared to literature quasars ($5.0\leq z \leq7.5$) and observed stellar M-, L-, and T-dwarfs\protect\footnotemark.}
     \label{fig:selection_colors}
\end{figure}

After creating the initial sample, we apply optical and near-infrared (NIR) colour cuts, which are similar to those used in other high-$z$ quasar searches (see e.g. \citealt{Banados2016ApJS..227...11B, Jiang2016ApJ...833..222J, Matsuoka2018ApJS..237....5M}). The wavelength range between the \textit{r}- and \textit{z}-band allows for probing the Lyman break of quasars in the redshift range of $4.9 \lesssim z \lesssim 6.6$. Therefore, we selected sources with \textit{r}$-$\textit{z} > 1.4 and required a non-detection in the \textit{g}-band with a signal-to-noise ratio (S/N) below 3. In addition, to remove stellar contaminants we included a selection criterion redward of the Lyman break of $-0.5< z-W1 < 2.5$ (see Fig.~\ref{fig:selection_colors}). These selection criteria are sensitive to quasars over a broad redshift range, including the ``quasar redshift gap'' at $z\sim$5.3-5.7 created by conventional colour selections between the $r$ and $i$-band (see e.g. \citealt{Yang2019ApJ...871..199Y}). Furthermore, our loose colour selection allows for the possibility of finding overlooked quasars outside of the traditional colour regions. Our optical and NIR selection criteria can be summarised as follows: 

\begin{align}
    \text{S/N} \ \textit{g} &< 3 \\
    r - z &> 1.4 \\
    -0.5< z-W1 &< 2.5  \\
    \notag\quad (\text{if S/N W1} &> 3).
\end{align}

\footnotetext{\url{https://docs.google.com/spreadsheets/d/1i98ft8g5mzPp2DNno0kcz4B9nzMxdpyz5UquAVhz-U8/edit?usp=sharing}}

\subsection{LoTSS-DR2}
\label{subsec:lotss}
For our radio selection, we use the LoTSS-DR2 144 MHz survey (see \citealt{Shimwell2022A&A...659A...1S}), which contains a total of $\sim$4.4 million radio sources split over two fields. These are denoted as the 0hr and 13hr fields and cover sky areas of 1457 and 4178 deg$^2$, with median rms values of 106 and 74 $\mu$Jy beam$^{-1}$, respectively. 

The next step in our selection process was cross-matching with the $>5\sigma$ detected radio sources in the LoTSS-DR2 radio catalogues. The LoTSS-DR2 resolution is 6$\arcsec$ with an astrometric accuracy of $\sim$0.2$\arcsec$ at $\geq$ 20 mJy, and reaching an accuracy of 0.5$\arcsec$ at an S/N of 5. By cross-matching LoTSS-DR2 with the quasar catalogue from the Sloan Digital Sky Survey (SDSS; \citealt{Paris2018A&A...613A..51P}) at $0\leq z \leq5$  and with known high-$z$ quasars \citep{Ross2020MNRAS.494..789R}, a search radius of 2$\arcsec$ was found to be sufficiently large to identify their radio counterparts without being dominated by false associations. Before cross-matching we have $\sim$1.6 million candidates in the LoTSS-DR2 region. Cross-matching the high-$z$ quasar candidates within a radius of 2$\arcsec$ using the Legacy $z$-band and LoTSS-DR2 positions, drastically reduced our candidate sample to 310 potential high-$z$ quasars, including 11 known quasars with $z > 5$ from \cite{Ross2020MNRAS.494..789R}. The radio-optical source separation is further discussed in Section~\ref{subsec:optical_radio_sep}. We note that we do not apply a radio compactness criterion to allow for marginally resolved radio emission from larger extended radio sources in the sample (see e.g. \citealt{Miley1971MNRAS.152..477M, Miley1968Natur.218..933M, Carilli1997ApJS..109....1C, saxena2019MNRAS.489.5053S}) and also accounting for aspects such as variable ionospheric smearing and incomplete deconvolution at low flux densities which can artificially enhance the source sizes (see \citealt{Shimwell2022A&A...659A...1S}).

\subsection{SED fitting}
\label{subsec:sedfitting}

We carried out SED fitting on the sample of 310 potential high-$z$ quasars using the photometric redshift code \textsc{Eazy} \citep{brammer2011ApJ...739...24B}, to further constrain their redshifts. Besides the Legacy Surveys photometry, we include forced photometric measurements of PS1 images in the \textit{g,r,i,z,y} bands (see  \citealt{Chambers2016arXiv161205560C}) and UKIDSS Hemisphere Survey $J$-band images (UHS; \citealt{Dye2018MNRAS.473.5113D}) within a circular aperture (of 1.5$\arcsec$) at the Legacy $z$-band source position. The majority of our candidates (72\% or 223/310) are detected in the PS1 catalogue, but including upper limits also for the non-detected sources below a 5$\sigma$ significance contributes to constraining the SED shape. In the UHS J-band 33\% (103/310) of our candidates are detected at $\geq5\sigma$ significance in the catalogues, and 16\% are not covered by UHS.

For the SED fitting, we used galaxy and quasar templates generated by \cite{Brown2014ApJS..212...18B} and \cite{Brown2019MNRAS.489.3351B}, respectively, which span a broad range of galaxy and quasar types and include additional reddening to take the full range of spectral slopes into account.
From the resulting $\chi^2$ redshift distributions, we required integrated photo-$z$ probabilities of at least 50\% above $z>5$ and $z_{\text{phot}} - 2 \times z_{\text{phot,err}} > 4.5$ for each candidate, which yielded a total of 253 high-$z$ candidates.

Finally, we visually inspected all remaining candidates to remove any spurious sources and sources with wrongly associated radio detections (e.g. cosmic rays and association with neighbouring optical sources), which resulted in a final sample of 142 candidates for spectroscopic follow-up observations. 

All steps taken in the selection procedure are summarised in Table ~\ref{tab:selection}. The requirement of a radio detection drastically decreases the number of candidates from 1,573,757 to 310, which makes the final sample tractable for follow-up spectroscopy.

\begin{table}
\caption{Overview of selection steps taken to create our candidate target list for spectroscopic follow-up.}
\label{tab:selection}      
\centering
\resizebox{\columnwidth}{!}{
\begin{tabular}{c c c c} 
\hline\hline
Step & Description & Num. sources & Num. sources \\ 
 & & remaining & removed \\ 
\hline
1 &  Num. sources in catalogue  & 2,434,158 & \\
 &  in LoTSS-DR2 footprint  &  & \\
 & with flag criteria: &  & \\
 &  $z_{\text{phot}}>3.5$  &  & \\
  &\texttt{fracflux} \textunderscore \texttt{r/z} < 0.2  & & \\
  & \texttt{maskbits=0} & & \\  

2 & Colour cuts: & 1,573,757 & 860,401 \\
  & S/N \ \textit{g} < 3  & & \\
  &  $r - z$ > 1.4 & & \\
  &  -0.5< $z-W1$ < 2.5 & & \\
  & (if S/N $W1$ > 3) & & \\
3 & Cross-match LoTSS-DR2 & 310 & 1,573,447 \\
4 & EAZY SED fitting & 253 & 57 \\
5 & Visual inspection & 142 & 111 \\
\hline \hline
\end{tabular}
} 
\end{table}

\section{Spectroscopic observations}
\label{sec:spectroscopic_obs}

To confirm our high-$z$ candidates we have conducted spectroscopic observations using primarily the Faint Object Camera and Spectrograph (FOCAS; \citealt{Kashikawa2002PASJ...54..819K}) on the Subaru Telescope on Mauna Kea, Hawaii and LRS2 \citep{Chonis2016SPIE.9908E..4CC} on the Hobby-Eberly Telescope at the McDonald observatory in Texas, USA (HET; \citealt{Ramsey1998AAS...193.1007R, Hill2021AJ....162..298H}). In addition, several of our targets were observed using the Low Resolution Imaging Spectrometer (LRIS; \citealt{Oke1995PASP..107..375O}) on Keck telescope on Mauna Kea.
The observational details of our newly discovered high-$z$ quasars are summarised in Table~\ref{tab:spec_obs}. An additional 2 candidates were followed-up with the FOcal Reducer/low dispersion Spectrograph 2 (FORS2; \citealt{Appenzeller1992Msngr..67...18A}) on the Very Large Telescope (VLT) at Paranal Observatory in Chile, but these were found to be low-redshift contaminants and therefore do not appear in Table~\ref{tab:spec_obs}.

The Subaru FOCAS observations were conducted using the VPH850 grating and SO58 filter ($5800 - 10350$ \r{A}) with a slit width of 1$\arcsec$ and spectral resolution of $R\sim600$. The spectra were reduced following the standard procedure\footnote{\url{https://subarutelescope.org/Observing/DataReduction/Cookbooks/FOCAS_cookbook_2010jan05.pdf}} using Image Reduction and Analysis Facility (IRAF; \citealt{Tody1986SPIE..627..733T}), which includes bias subtractions, flat fielding, distortion corrections, wavelength calibration and sky subtraction. Cosmic rays were removed using the algorithm \textsc{L.A.Cosmic} based on the Laplacian edge detection algorithm of \cite{vanDokkum2001PASP..113.1420V}. The spectra were flux calibrated using standard stars HZ4, Feige34, and Feige110 observed on August 14th, November 25th, and December 27th 2021, respectively.

The HET observations reported here were obtained with the LRS2 integral field spectrograph. LRS2 comprises two double spectrographs each with a 6$\arcsec\times$12$\arcsec$ integral field input separated by 100$\arcsec$ on sky: LRS2-B (3650 - 6950 \r{A}) and LRS2-R (6450 - 10500 \r{A}). The integral field units have 100\% fill-factor. There are two channels for each spectrograph: UV and "Orange" for LRS2-B and "Red" and "FarRed" for LRS2-R. We observed our targets with the LRS2-R spectrograph with exposure times of 1500-1800s. The HET observations were reduced using the HET LRS2 pipeline Panacea\footnote{\url{https://github.com/grzeimann/Panacea}} to perform the initial reductions including fiber extraction, wavelength calibration, astrometry, sky subtraction, and flux calibration. For the initial sky subtraction for each channel a biweight average \citep{Beers1990AJ....100...32B} of the fibers excluding the target was used. Fringing and varying spectral resolution cause systematic residuals in the initial sky subtraction. We use a principle component analysis similar to the algorithm of Zurich Atmosphere Purge (ZAP; \citealt{Soto2016MNRAS.458.3210S}) to model the sky subtraction residuals. After the final sky subtraction, we combine fiber spectra from the two channels, Red and FarRed, into a single data cube accounting for differential atmospheric refraction.  We then identified the target quasar in each cube and extract our target spectra in a 1.5$\arcsec$ aperture. ILTJ0037+2410 has been observed with both HET LRS2 and Subaru FOCAS and can therefore be used to compare the resulting spectra. This comparison will be discussed in Section~\ref{subsec:comparison_sub_het}.

The Keck LRIS spectra were reduced using the data reduction package PypeIt \citep{Prochaska2020JOSS....5.2308P}. After basic calibrations (e.g. bias calibration, wavelength calibration), the object extraction \citep{Horne1986PASP...98..609H} and b-spline \citep{Kelson2003PASP..115..688K} sky subtraction were performed together. All individual 1D spectra were flux-calibrated with the standard star GD153.

\begin{figure}
\centering
   \includegraphics[width=\columnwidth, trim={0.0cm 0cm 0cm 0.0cm}, clip]{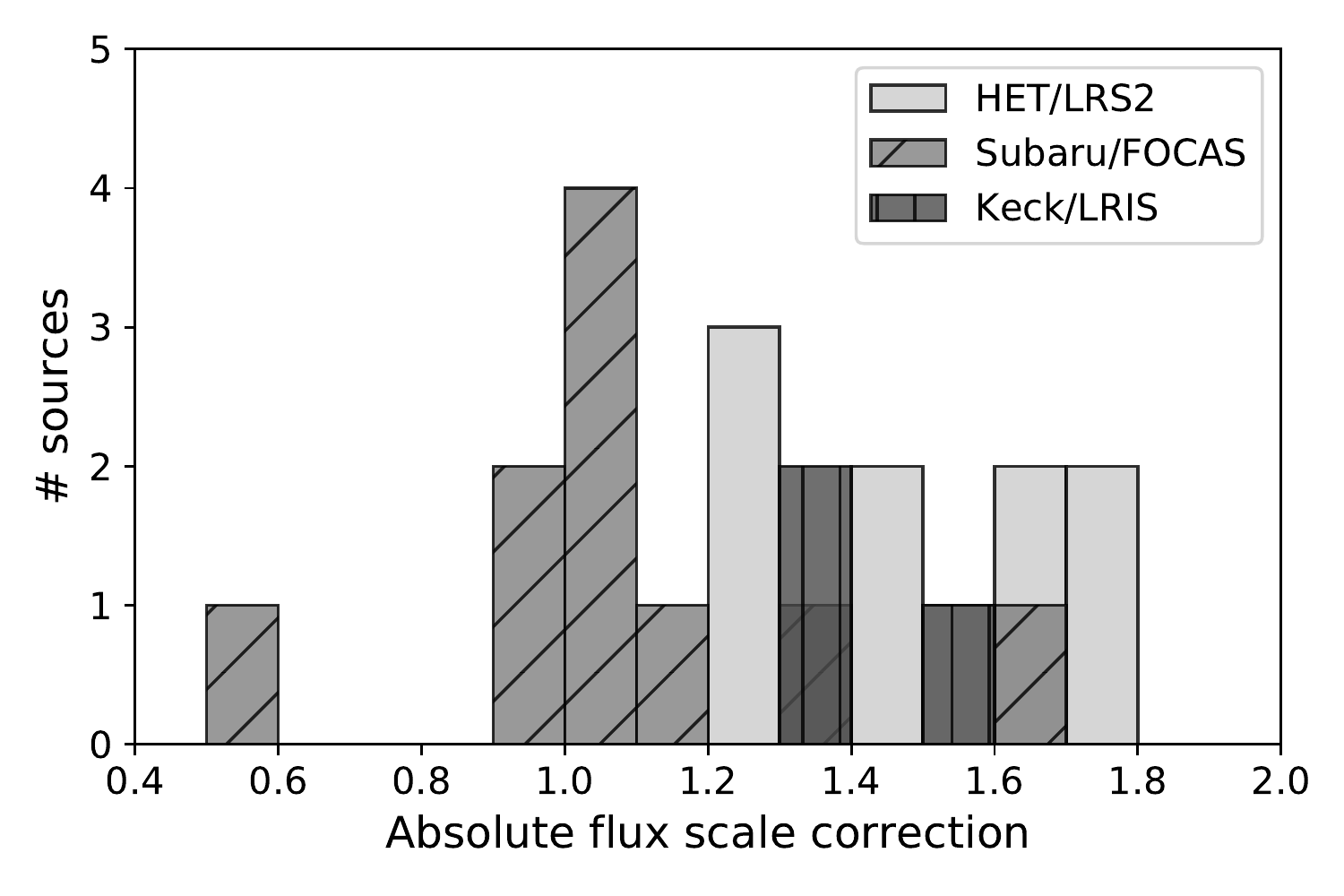}
     \caption{Absolute flux scale correction factors between the quasar spectra and PS1 $i$-band and Legacy $z$-band photometry. A correction factor of $>1$ means the spectrum continuum is lower than the photometric measurement. These corrections do not affect the redshift and Ly$\alpha$ EW estimates, but do decrease or increase the measured Ly$\alpha$ emission line flux.}
     \label{fig:abs_flux_scale}
\end{figure}

To investigate the absolute flux scale accuracy of the reduced spectra, we convolved the spectra with the PS1 and Legacy Surveys photometric filter transmission profiles (including atmosphere) above the Lyman break and compare this to the photometric catalogue fluxes. The result is shown in Fig.~\ref{fig:abs_flux_scale} and the values for all the discovered quasars range between 0.58-1.78. These values were determined using the PS1 $i$-band if there was a photometric detection. Otherwise, the Legacy Surveys $z$-band values were used. The spectra are expected to result in lower fluxes than the photometry (absolute flux scale corrections $>1$) due to slit losses for the Subaru/FOCAS and Keck/LRIS and losses due extraction of the quasars from the IFU for HET/LRS2. However, the quasar ILTJ2336+1842 at $z=6.6$ has a correction factor of 0.58, which is likely due to the Lyman break being close to red end of the Legacy $z$-band and telluric lines heavily affecting that part of the spectrum. The other two quasars below 1 deviate only slightly with correction factors of 0.97 and 0.98. The deviations in absolute flux scale do not affect the redshift determination or the Ly$\alpha$ EW estimates of the quasars. However, they do influence the measured Ly$\alpha$ emission line fluxes and were therefore used in error propagation (see Section~\ref{subsec:lya_method}). Hence, we do not use the correction factors to calibrate the spectra, but do include them in the error estimation.

\begin{table*}
\caption{Details of spectroscopic observations conducted of the newly discovered high-$z$ quasars (sorted on RA). }
\label{tab:spec_obs}      
\centering
\resizebox{0.6\textwidth}{!}{
\begin{tabular}{c c c c c}  
\hline\hline
Quasar & Date obs. & Telescope & Instrument & Exposure time \\ 
\hline
ILTJ0026+2516 & 27-12-2021 & Subaru & FOCAS & 900 s \\
ILTJ0037+2410 & 14-08-2021 & Subaru & FOCAS & 600 s \\ 
 & 09-09-2021 & HET & LRS2 & 1800 s \\ 
ILTJ0045+2749 & 14-08-2021 & Subaru & FOCAS & 2x900 s \\  
ILTJ0121+2940 & 25-11-2021 & Subaru & FOCAS & 900 s \\
ILTJ0912+6658 & 25-11-2021 & Subaru & FOCAS & 600 s \\
ILTJ0922+4815 & 25-11-2021 & Subaru & FOCAS & 600 s \\
ILTJ0952+6311 & 10-11-2021 & HET & LRS2 & 1800 s \\ 
ILTJ1013+3518 & 02-11-2021 & HET & LRS2 & 1800 s \\ 
ILTJ1037+4033 & 11-11-2020 & HET & LRS2 & 3x1500 s \\ 
ILTJ1043+4048 & 25-11-2021 & Subaru & FOCAS & 300 s \\
ILTJ1054+4553 & 11-11-2021 & HET & LRS2 & 1800 s \\  
ILTJ1104+6631 & 27-01-2022 & HET & LRS2 & 1800 s \\  
ILTJ1133+4814 & 27-12-2021 & Subaru & FOCAS & 900 s \\
ILTJ1330+6224 & 30-12-2021 & HET & LRS2 & 1800 s \\
ILTJ1334+4750 & 28-03-2022 & HET & LRS2 & 1800 s \\
ILTJ1350+3748 & 27-04-2017 & Keck & LRIS & 2x600 s \\
ILTJ1401+4542 & 22-04-2022 & Keck & LRIS & 300 s \\
ILTJ1419+5718 & 30-12-2021 & HET & LRS2 & 1800 s \\
ILTJ1523+2935 & 22-04-2022 & Keck & LRIS & 300 s \\
ILTJ1601+3102 & 26-08-2021 & HET & LRS2 & 1800 s \\ 
ILTJ1650+5457 & 28-08-2021 & HET & LRS2 & 1500 s \\ 
ILTJ2201+2338 & 14-08-2021 & Subaru & FOCAS & 2x600 s\\
ILTJ2327+2454 & 07-05-2021 & HET & LRS2 &  1800 s\\ 
ILTJ2336+1842 & 27-12-2021 & Subaru & FOCAS & 900 s \\

\hline \hline
\end{tabular}
} 
\\ \vspace{0.1cm} \raggedright { \textbf{Notes.} The Subaru/FOCAS observations have taken place on 14th August 2021 (half night), 25th November 2021 (half night), and 27th December 2021 (full night) under proposal ID S21B-003 (PI: Y. Harikane). The HET/LRS2 observations have been conducted in service mode using LRS2 guaranteed time and awarded time under proposal ID UT22-1-008 and UT22-2-009 (PI: G. J. Hill).}
\end{table*}

\section{New quasars}
\label{sec:new_quasars}

We have performed spectroscopic follow-up observations of 80 candidates in our sample ($\sim$56\%), which has resulted in the discovery of 20 new high-$z$ quasars and the independent confirmation of 4 (which will be discussed below). Taking account of the 10 candidates that were not detected due to bad weather conditions, our success rate of spectroscopically confirming high-$z$ quasars from our candidate sample was 34\% (24/70). The contaminants we find are listed and discussed in Appendix \ref{sec:appendix_contaminants}. We note that four of our quasars have been previously published in other work, but have been independently discovered in this work. These are ILTJ1419+5718 \citep{McGreeg2018AJ....155..131M}, ILTJ1523+2935 \citep{Wenzl2021AJ....162...72W}, ILTJ1043+4048 and ILTJ1334+4750  \citep{Lyke2020ApJS..250....8L}. Furthermore, some of our quasar candidates have been observed based on earlier catalogues and selection criteria (see Section \ref{subsec:notes_individual} for details). Our final selection described in Section \ref{sec:candidate_selection} includes 22 new quasars and 42 contaminants, yielding an identical success rate of $\sim$34\% (22/64). This high success rate can be attributed to the radio selection by decreasing the number of stellar contaminants in the sample.

\begin{figure*}
\centering
   \includegraphics[ width=0.98\textwidth, trim={3.5cm 6cm 3cm 6.0cm}, clip]{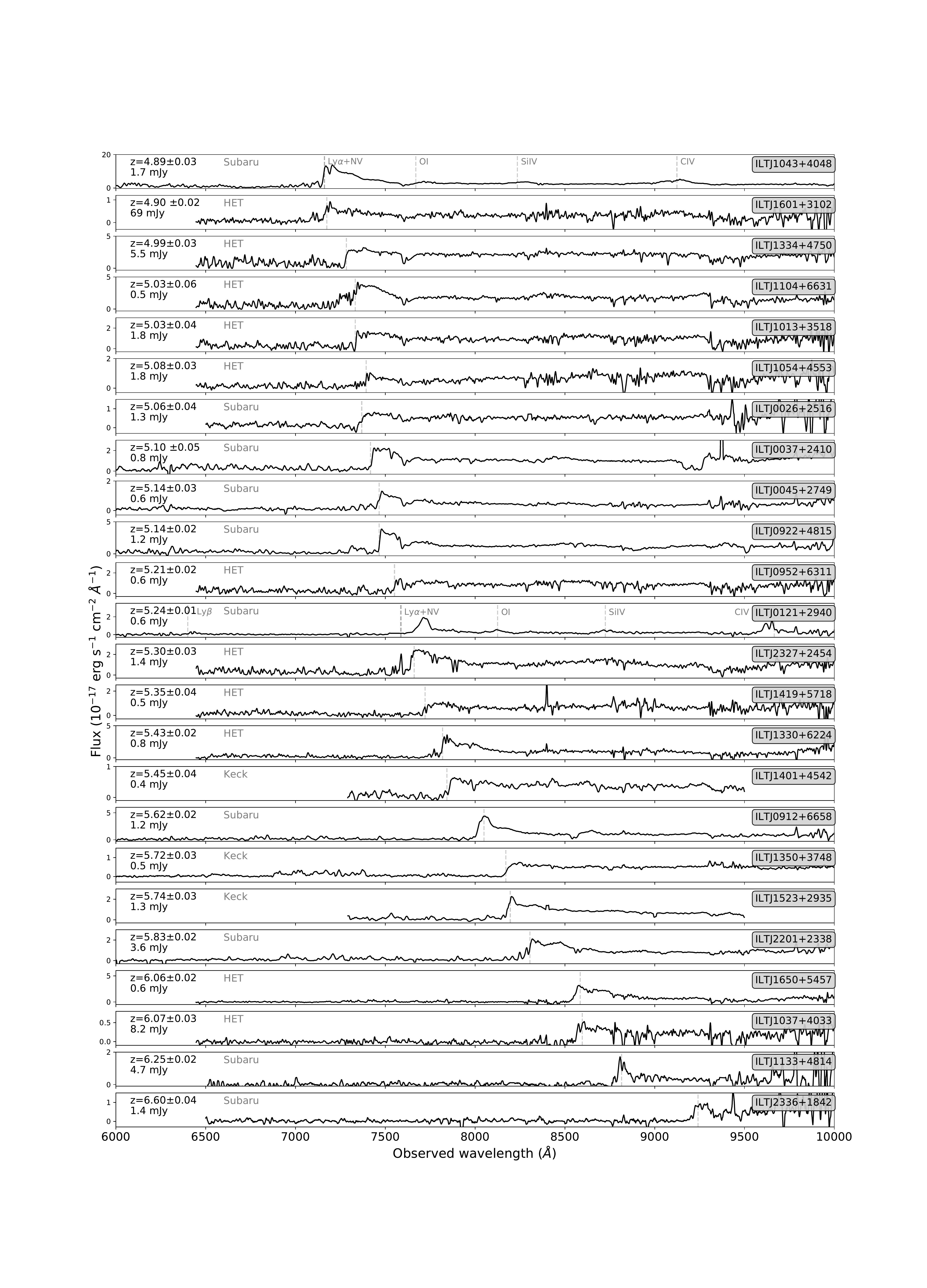}
     \caption{Observed-frame optical spectra obtained with Subaru/FOCAS, HET/LRS2, and Keck/LRIS of the newly confirmed quasars sorted on spectroscopic redshift. The total flux measurements at 144 MHz are also indicated on the left hand side of each spectrum. The detected emission lines are indicated by the grey dashed lines, which are Ly$\alpha$+N\textsc{v} emission lines unless indicated otherwise.}
     \label{fig:spectra}
\end{figure*}

The spectra of the 24 newly discovered quasars are shown in Fig.~\ref{fig:spectra}. Their redshifts range from $4.9$ to $6.6$ (see Section~\ref{subsec:redshifts}). The discovery of quasars in this large redshift range was possible due to the wavelength gap between the Legacy survey's $r$ and $z$-band filter. The Legacy, WISE, and LOFAR images of all new quasars are shown in Appendix \ref{sec:appendix_cutouts} and quasar properties are summarised in Table \ref{tab:quasar_properties}. The optical spectra and measured properties are provided as a data product \footnote{\label{cdstable}The optical spectra and quasar properties are available in electronic form
at the CDS via anonymous ftp to cdsarc.u-strasbg.fr (130.79.128.5)
or via \url{http://cdsweb.u-strasbg.fr/cgi-bin/qcat?J/A+A/}. For now the data can be found here: \url{https://doi.org/10.5281/zenodo.7142693}}.  

\subsection{Redshifts}
\label{subsec:redshifts}
Determining accurate redshifts for high-$z$ quasars at optical wavelengths is challenging, because the most prominent Ly$\alpha$ emission line is strongly affected by absorption. More accurate redshift estimates can be made by using atomic and molecular emission lines from the quasar, such as [CII], MgII or CO emission lines (e.g. \citealt{Wang2010ApJ...714..699W, Eilers2020ApJ...900...37E, Schindler2020ApJ...905...51S}). However, the limited wavelength range of our observations does not include any of these emission lines. Therefore, in this work, we use a template fitting method of composite high-$z$ quasar spectra (similar to \citealt{Banados2016ApJS..227...11B}) to constrain the redshift using the Lyman break and redward continuum. We fitted each quasar using the three composite quasar templates constructed by \cite{Banados2016ApJS..227...11B} for weak, moderate, and strong Ly$\alpha$ emission and the composite quasar template constructed by \cite{Selsing2016A&A...585A..87S}. Typically, the redshift estimates following from the $\chi^{2}$ minimisation of these templates agree well with maximum deviations of $\sim$0.05, but for our final redshift estimates we used the best fitting template. To determine errors on the spectroscopic redshift estimates, we convert the $\chi^{2}$ values as a function of redshift to a probability density distribution and fit the peak of the distribution using a Gaussian fit. The error on the redshift is then given by the standard deviation of the best Gaussian fit. This method yields errors on the $z_{\text{spec}}$ between 0.01-0.06 with a median of $0.03$ (see Table \ref{tab:quasar_properties}). We emphasise that follow-up (near-)infrared or sub-millimetre observations are necessary to obtain more precise systemic redshifts for these quasars. 

\subsection{Emission lines}
\label{subsec:lya_method}

The Ly$\alpha$ emission line at 1216 \r{A} is the most prominent line we detect in the optical spectra of our quasars. Additionally, for two quasars we also detect Ly$\beta$, O\textsc{i}, Si \textsc{iv}, and C\textsc{iv} emission lines (see Section~\ref{subsubsection:ILTJ0121+2940} and \ref{subsubsection:ILTJ1043+4048} for details). The strength of the Ly$\alpha$ emission line varies considerably between the different quasars (see Fig.~\ref{fig:spectra}) and also contains the contribution from the N\textsc{v} emission line at $\sim$1240 \r{A} and Si \textsc{ii} line at 1263 \r{A}, with N\textsc{v} being the most dominant. To investigate the emission line properties as a function of redshift and radio luminosity, we estimate the Ly$\alpha$+N\textsc{v} luminosity and equivalent width (EW) using the method of \cite{Diamond-Stanic2009ApJ...699..782D}. This method entails fitting a power-law of the form $f_{\lambda} = C \times \lambda^{\beta}$ to the continuum between the rest-frame wavelength regions uncontaminanted by emission lines in the range of [1285, 1295], [1315, 1325], [1340, 1375], [1425, 1470], and [1680, 1710] \r{A}. However, note that the final wavelength range of [1680, 1710] is not covered by all our quasars above $z>5$. We fix the slope of the power-law fit $\beta$ to the values -1.5 and -1.7 (as done by \citealt{Banados2016ApJS..227...11B}) to avoid unphysical values of the continuum slope caused by noise and telluric contamination. We take into account the error on the spectroscopic redshift by fitting the power-law 10000 times to the rest-frame spectrum determined by redshifts drawn from a normal distribution around the $z_{\text{spec}}$. The Ly$\alpha$+N\textsc{v} fluxes and EWs are determined by integrating between 1160-1290 \r{A} above the continuum using the power-law fits (and averaging over the $\beta=$-1.5 and -1.7 solutions). For each source, the final estimated Ly$\alpha$ flux and EW are given by the 50th percentile on the resulting distributions and the errors by the 16th and 84th percentiles.  
As discussed in Section~\ref{sec:spectroscopic_obs}, the measured line fluxes are potentially affected by the inaccurate absolute flux scale corrections of the spectra, therefore the correction factors from Fig.~\ref{fig:abs_flux_scale} have been also been included in the errors on the Ly$\alpha$+N\textsc{v} flux.

\subsection{Rest-frame UV magnitudes}
To determine the UV rest-frame magnitudes of our quasars we use a similar SED fitting procedure on their photometry as described in Section~\ref{subsec:sedfitting}, but fix the templates to the spectroscopic redshift. From the best fits we derive the UV rest-frame magnitudes at 1450, 2500, and 4400 \r{A} by measuring the average flux with tophat filters centred on the respective central wavelength with a width of 100 \r{A}. To obtain the final UV magnitudes and their uncertainties we adopt a simple Monte Carlo method, where we duplicate each source 500 times and randomly adjust the fluxes by drawing from a normal distribution around each catalogue flux value with the standard deviation determined by the catalogue flux errors. The UV magnitudes are then given by the 50th percentile values, and the errors by the 16th and 84th percentile of the resulting distribution. We note that the intrinsic UV magnitudes are brighter than the observed ones due to dust reddening (e.g. \citealt{Calistro2021A&A...649A.102C}). The dust reddening of our quasars is discussed in Section~\ref{subsec:redquasars}.

\subsection{Notes on individual quasars}
\label{subsec:notes_individual}
Here we discuss some of our newly discovered quasars that have notable properties, such as detected emission lines in addition to Ly$\alpha$, and quasars that have not been selected by our main selection method as described in Section~\ref{sec:candidate_selection}. 

\subsubsection{ILTJ0037+2410}
\label{subsec:comparison_sub_het}
The quasar ILTJ0037+2410 was observed with both by HET and Subaru, allowing a comparison of the absolute flux scale and the wavelength calibration. Both the derived Ly$\alpha$+N\textsc{v} luminosities and the measured EW$_{\text{Ly}\alpha + \text{N}\textsc{v}}$ are in good agreement with luminosities of 10$^{43.68^{+0.05}_{-0.14}}$ and 10$^{43.69^{+0.16}_{-0.17}}$ erg s$^{-1}$ and EWs of 14$^{+2}_{-6}$ and 16$^{+2}_{-3}$ for the Subaru and HET spectrum, respectively. The spectroscopic redshifts also agree very well with a difference of $\sim$0.01 from our fitting procedure (see Section~\ref{sec:spectroscopic_obs}). 

\subsubsection{ILTJ1013+3518}
This quasar was not selected by the method described in Section~\ref{sec:candidate_selection}, because it was omitted from the catalogues of \cite{Duncan2022MNRAS.512.3662D} due to its unreliable photometry caused by contamination of a neighbouring object. However, it was selected and observed prior to the creation of this final sample and confirmed to have a redshift of $z=5.03$. 

\subsubsection{ILTJ1350+3748}
This quasar has been identified as a dropout and confirmed using Keck/LRIS in 2017 (see also Banados et al. in prep, who independently discovered it).

\subsubsection{ILTJ0121+2940} 
\label{subsubsection:ILTJ0121+2940}
This quasar and ILTJ1043+4048 are the only quasars where we have detected additional emission lines. In the spectrum of ILTJ0121+2940 we have identified as Ly$\beta$, O\textsc{i}, Si \textsc{iv}, and C\textsc{iv}, which can also be used to determine the spectroscopic redshift. A redshift of $z=5.24\pm0.01$ would match the Ly$\beta$, O\textsc{i}, Si \textsc{iv} emission lines with a blueshift of C\textsc{iv}. However, this redshift does not match the Ly$\alpha$ line at $\sim$7700 \r{A}, which would be shifted to $\sim$7585 \r{A} for $z=5.24$ and our template fitting procedure yielded a redshift of $z=5.34\pm0.02$ instead (see Section \ref{subsec:redshifts}). Coincidentally, this wavelength region (7529-7600 \r{A}) had to be masked out, due to a cosmic ray at that exact location. Therefore, the redshift of this quasar remains unclear and both redshift estimates are quoted in Table ~\ref{tab:quasar_properties}. For subsequent plots we adopt the redshift determined by the Ly$\beta$, O\textsc{i}, Si \textsc{iv} emission lines of $z=5.24\pm0.01$. We note that potentially strong Ly$\alpha$ absorption cannot be explained by an IGM damping wing, since the quasar is at $z<5.5$ (see e.g. \citealt{Wolfe2005ARA&A..43..861W}), nor by an absorber near the quasar, because of the presence of Ly$\beta$ emission (e.g. \citealt{Banados2019ApJ...885...59B}). This quasar could potentially be a broad absorption line (BAL) quasar with narrow Ly$\alpha$ (e.g. \citealt{Ross2015MNRAS.453.3932R}) or extremely high velocity outflows (e.g. \citealt{Rodriguez2020ApJ...896..151R}) causing Ly$\alpha$ and N\textsc{v} absorption.  

The detection of the C\textsc{iv}$\lambda\lambda$1548.2,1550.8 doublet allows for estimating the SMBH mass by measuring the full width at half maximum (FWHM) and continuum luminosity as often used to derive BH masses at high-$z$ (see e.g. \citealt{Vestergaard2006ApJ...641..689V, Kelly2010ApJ...719.1315K, Park2013ApJ...770...87P}). To determine the continuum around the C\textsc{iv} line, we fit a power-law to the rest-frame spectrum using the same method as described in Section ~\ref{subsec:lya_method}. Subsequently, the continuum luminosity is calculated using the power-law fit between 1500-1600 \r{A} and the FWHM of C\textsc{iv} is determined by fitting a Gaussian profile to the continuum subtracted spectrum. The virial BH mass is finally calculated by
\begin{align}
    \log\Big(\frac{M_{BH}}{M_{\odot}}\Big) = a + b\log\Big(\frac{L_{\text{cont}}}{10^{44} \ \text{erg} \ \text{s}^{-1}}\Big) + c\log\Big( \frac{\text{FWHM}}{\text{km} \ \text{s}^{-1}} \Big),
\end{align}
with (a, b, c) = (0.66, 0.53, 2.0) as determined by \cite{Zuo2020ApJ...896...40Z} using the calibrations from \cite{Vestergaard2006ApJ...641..689V}. This results in an estimated BH mass for ILTJ0121+2940 of $\log\Big(\frac{M_{BH}}{M_{\odot}}\Big) = 8.58^{+0.13}_{-0.03}$, which is in line with previous high-$z$ SMBH masses found by e.g. \cite{Fan2001AJ....122.2833F} and \cite{Jiang2007AJ....134.1150J}. The errors on this BH mass estimate include the absolute flux scale uncertainties (see Fig.~\ref{fig:abs_flux_scale}) and the redshift uncertainties, which both influence the continuum luminosity and subsequently the derived BH mass.

Note that this BH mass might be overestimated due to blueshifting of the C \textsc{iv} emission line, of $\sim10,000$ km s$^{-1}$ for $z=5.24$ assuming equal contribution of both components of the C \textsc{iv} with a rest-frame wavelength of 1549.98 \r{A} (see \citealt{Coatman2016MNRAS.461..647C}). This blueshifting is likely caused by strong outflows and a high Eddington luminosity ratio (see e.g. \citealt{Coatman2016MNRAS.461..647C,Marziani2019A&A...627A..88M,Zuo2020ApJ...896...40Z}).

\subsubsection{ILTJ1043+4048}
\label{subsubsection:ILTJ1043+4048}
Similarly to ILTJ1013+3518, this quasar was omitted from the final sample due to a $g$-band S/N of $>5 \sigma$ in the Legacy Surveys DR8 catalogue used in \cite{Duncan2022MNRAS.512.3662D}. However, in the Legacy Surveys DR7 catalogue the quasar has a negative $g$-band S/N of $-4.9$ and was therefore previously selected for follow-up observations. 

Also for this quasar we identify Ly$\beta$, O\textsc{i}, Si \textsc{iv}, and C\textsc{iv} in the spectrum. In this case the observed wavelengths of the lines are consistent with the redshift determined by template fitting of $z=4.89\pm0.03$. Performing the same fitting procedure of the C\textsc{iv} line yields supermassive black hole mass of $\log\Big(\frac{M_{BH}}{M_{\odot}}\Big) = 9.84^{+0.02}_{-0.01}$, which is slightly higher than expected from its absolute UV magnitude ($\log\Big(\frac{M_{BH}}{M_{\odot}}\Big) = 9.1$) when assuming a constant Eddington ratio and bolometric correction \citep{Inayoshi2020ARA&A..58...27I}. However, similar masses have been reported for other quasars around $z\sim5$, but it is at the high end of the expected SMBH mass range (see e.g. \citealt{Vestergaard2008ApJ...674L...1V, Wang2015ApJ...807L...9W, Wu2015Natur.518..512W, Sbarrato2021A&A...655A..95S}). Observations of the Mg\,\textsc{ii} line could provide a more accurate measurement of the SMBH mass of this source, due to the effects of outflows on the C\textsc{iv} emission line (e.g. \citealt{Coatman2016MNRAS.461..647C, Onoue2019ApJ...880...77O, Shen2019ApJ...873...35S}).  

\begin{figure*}
\centering
   \includegraphics[width=0.75\textwidth, trim={0.0cm 0.3cm 0cm 0.0cm}, clip]{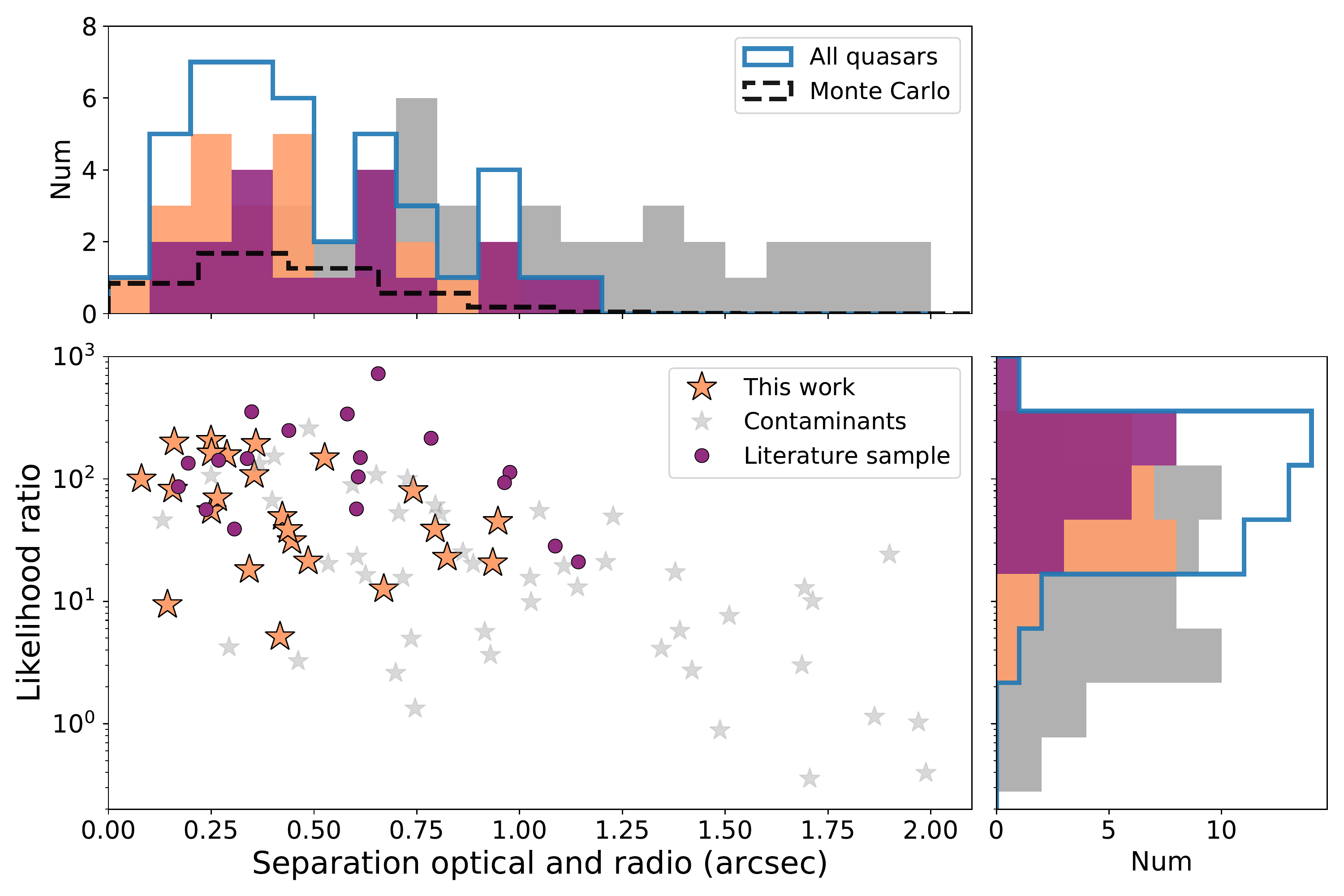}
     \caption{Association likelihood ratios versus optical-radio separations of our observed sample. The blue histograms show the distribution for the quasar sample from this work and literature combined. We demonstrate that a similar distribution (p-value = 0.5) of separations can be created by a simple Monte Carlo simulation using the positional offsets from both the Legacy and LoTSS-DR2 catalogue. From this distribution we derive the chance of a quasar having an optical-radio separation larger than 1$\arcsec$ to be $\sim$2\% when assuming no intrinsic difference in the position of the radio and optical emission.} 
     \label{fig:likelihoodratio}
\end{figure*}

\subsection{Optical-radio separation}
\label{subsec:optical_radio_sep}

Our observing campaign has been successful in selecting high-$z$ quasars with a success rate of 34\%. However, we could have further improved this by lowering the separation limit between the optical and radio positions. The separation of the quasars and contaminants as a function of the association likelihood ratio (LR; see e.g. \citealt{Williams2019A&A...622A...2W}) are shown in Fig.~\ref{fig:likelihoodratio}, which clearly shows a downward trend in LR with separation, as expected. These LRs have been calculated for LoTSS-DR2 to identify the most likely optical counterparts in the Legacy Surveys for each non-complex detected radio source (see Hardcastle et al. in prep.). In short, the LR is defined as the ratio between the probability of an object being the true counterpart and being a random interloper. To calculate the LRs for optical counterparts in LoTSS-DR2 a maximum likelihood method is used (based on \citealt{Williams2019A&A...622A...2W}), which uses a prior probability that a radio source has an optical counterpart with a certain optical colour and magnitude, the surface density of sources with specific colours and magnitudes, and the probability distribution function for the offset between the radio and optical sources. To avoid biases due to galaxy clustering, the fields around each radio source are compared to other random sky positions. Source types are not considered in the LR calculation. For more details on the LRs we refer to \cite{Williams2019A&A...622A...2W} and Hardcastle et al. in prep.

In the blue histogram in the top panel of Fig.~\ref{fig:likelihoodratio} the distribution of separations is shown for the quasar sample from this work combined with the quasar literature sample. We demonstrate that a similar distribution can be created by a simple Monte Carlo simulation using the positional offsets from both the Legacy and LoTSS-DR2 catalogue. The average astrometric precision of the LOFAR sources in the LoTSS-DR2 region is $\sigma_{\text{RA}} = 0.22\arcsec$ and $\sigma_{\text{DEC}} = 0.20\arcsec$ for S/N$>20$ and the precision decreases gradually for lower S/N up to a maximum of 0.5$\arcsec$ for an S/N of $\sim$5 \citep{Shimwell2022A&A...659A...1S}. The optical positions in the Legacy Surveys have a high (median) astrometric accuracy of $\sim$0.02\arcsec \citep{dey2019AJ....157..168D}, therefore the positional uncertainties are dominated by the LOFAR astrometric offsets and by the fact that some of these have extended jet emission that can offset the radio centroid from the true core. The positional uncertainties for our quasar sample and literature sample are on average $\sigma_{\text{RA}} = 0.40\arcsec$ and $\sigma_{\text{Dec}} = 0.29\arcsec$. The Monte Carlo distribution created using these positional uncertainties is statistically similar to the quasar distribution with a p-value of 0.5. From this distribution we derive the chance of a quasar having an optical-radio separation larger than 1$\arcsec$ to be $\sim$2\% and larger than 1.2$\arcsec$ to be less than 1\%. This is in agreement with all newly discovered and known quasars having optical-radio separation offsets smaller than $1.2\arcsec$. Therefore taking this smaller offset ($<1.2\arcsec$) as a selection criterion of our sample, the success rate increases to 42\%. The increased contamination fraction with separation indicates that there might be false associations amongst the contaminants and shows that quasars at these redshifts typically do not have large offsets ($>1\arcsec$) in their radio emission. For our assumed cosmology, an offset of $1\arcsec$ corresponds to 5.7$\pm$2.9 kpc physical size at $z=6$ with the error determined by the astrometric accuracy of LOFAR. High-resolution radio observations are necessary to study the resolved jet structure (e.g. using the international LOFAR baselines; \citealt{Morabito2022A&A...658A...1M} or the Very Long Baseline Array (VLBA); \citealt{Momjian2018ApJ...861...86M}). 

The contaminants we find in this work are listed in Table \ref{tab:contaminants} and briefly discussed in Appendix \ref{sec:appendix_contaminants}. Regardless of the nature of the contaminants, for future high-$z$ quasars searches our results support the use of stricter optical-radio separation criteria to ensure maximum purity of the candidate sample. Based on the currently known radio-detected quasars, a separation of 1.2$\arcsec$ would be sufficient to retrieve all quasars in the LoTSS-DR2 survey.

\section{New quasar sample properties}
\label{sec:quasar_sample}

An overview of the observed properties of the newly discovered quasars is given in Table~\ref{tab:quasar_properties}. To put this new sample into the context of the previously known quasar population, we examine the radio and optical properties and compare them to the known high-$z$ quasars from literature work. For our literature sample, we primarily utilise the source list generated by \cite{Ross2020MNRAS.494..789R} with low-frequency radio properties determined by \cite{Gloudemans2021A&A...656A.137G} in the LoTSS-DR2 region, and include other recently published radio-loud quasars. 

\begin{table*}
\caption{Observed and measured physical properties of the discovered high-$z$ quasars.}
\label{tab:quasar_properties}      
\centering
\resizebox{0.98\textwidth}{!}{
\begin{tabular}{c c c c c c c c c c c c}
\hline\hline 
Quasar & Optical coordinates (J2000) & $z_{\text{spec}}$ & S$_{150\text{MHz}}$ & S$_{1.4\text{GHz,int}}$ & S$_{3\text{GHz,int}}$ & $\alpha_{1.4\text{GHz}}^{144\text{MHz}}$ & $\alpha_{3\text{GHz}}^{1.4\text{GHz}}$ & R & M$_{1450\text{\AA}}$ & EW$_{\text{Ly}\alpha+\text{N}\textsc{v}}$ & log$_{10}$ L$_{\text{Ly}\alpha+\text{N}\textsc{v}}$ \\ 
 & & & (mJy) & (mJy) & (mJy) & & & & & (\r{A})& (erg s$^{-1}$) \\
\hline 
ILTJ0026+2516  & 00:26:33.61 +25:16:53.12 & 5.06$\pm$0.04 & 1.3$\pm$0.3 & & & & & 42 $\pm$ 17 & -25.10$^{+0.14}_{-0.15}$ & 1.8 $^{+ 4.5 } _{- 1.5 }$ & 42.56 $^{+ 0.87 }_{- 0.39 }$  \\ 
ILTJ0037+2410  & 00:37:05.84 +24:10:53.30 & 5.10$\pm$0.05 & 0.8$\pm$0.2 &  & & & & 8 $\pm$ 3 & -26.08$^{+0.33}_{-0.04}$ & 14 $^{+ 2 } _{-6 }$ & 43.68 $^{+ 0.05 }_{- 0.14 }$ \\
ILTJ0045+2749  & 00:45:31.82 +27:49:44.28 & 5.14$\pm$0.03 & 0.6$\pm$0.1 &  & & & & 15 $\pm$ 6 & -24.85$^{+0.17}_{-0.12}$& 32 $^{+ 1 } _{- 1 }$ & 43.63 $^{+ 0.08 }_{- 0.08 }$ \\
ILTJ0121+2940  & 01:21:47.93 +29:40:32.89 & 5.24$\pm$0.01/5.34$\pm$0.02$^a$ & 0.6$\pm$0.2 &  & & & & 11 $\pm$ 4 & -24.27$^{+0.05}_{-0.42}$& 93 $^{+ 1 } _{- 1 }$ & 43.85 $^{+ 0.28 }_{- 0.28 }$ \\
ILTJ0912+6658  & 09:12:07.64 +66:58:46.95 & 5.62$\pm$0.02 & 1.2$\pm$0.1 &  & & & & 12 $\pm$ 3 & -26.43$^{+0.06}_{-0.03}$& 58 $^{+ 2 } _{-5 }$ & 44.32 $^{+ 0.03 }_{- 0.04 }$ \\
ILTJ0922+4815  & 09:22:03.57 +48:15:25.73 & 5.14$\pm$0.02 & 1.2$\pm$0.1 &  & & & & 11 $\pm$ 3 & -26.16$^{+0.03}_{-0.14}$ & 35 $^{+ 1 } _{- 1 }$ & 44.11 $^{+ 0.02 }_{- 0.02 }$ \\
ILTJ0952+6311  & 09:52:29.29 +63:11:37.83 & 5.21$\pm$0.02 & 0.6$\pm$0.1 &   & & & & 7 $\pm$ 3 & -26.22$^{+0.52}_{-0.04}$& 10 $^{+ 1 } _{- 1 }$ & 43.41 $^{+ 0.46 }_{- 0.46 }$ \\
ILTJ1013+3518  & 10:13:37.87 +35:18:49.81 & 5.03$\pm$0.04 & 1.8$\pm$0.1 & 1.12$\pm$0.13 & 0.68$\pm$0.11$^b$ & $-0.20\pm$0.06 & $-0.64\pm$0.26 & 31 $\pm$ 4 & -26.05$^{+0.11}_{-0.03}$ & 8.3 $^{+ 2.0 } _{- 1.5 }$ & 43.40 $^{+ 0.38 }_{- 0.38 }$ \\ 
ILTJ1037+4033  & 10:37:58.18 +40:33:28.74 & 6.07$\pm$0.03 & 8.2$\pm$0.2 & 9.42$\pm$0.15 & 8.39$\pm$0.29 & 0.06$\pm$0.01 & $-0.15\pm$0.05 & 1100 $\pm$ 300 & -25.25$^{+0.24}_{-0.08}$& 38 $^{+ 3 } _{- 12 }$ & 43.47 $^{+ 0.44 }_{- 0.45 }$ \\ 
ILTJ1043+4048  & 10:43:25.56 +40:48:49.45 & 4.89$\pm$0.03 & 1.7$\pm$0.2 & &  & & & 11 $\pm$ 3 & -26.23$^{+0.02}_{-0.02}$ & 95 $^{+ 5 } _{- 7 }$ & 44.85 $^{+ 0.04 }_{- 0.04 }$  \\ 
ILTJ1054+4553  & 10:54:04.02 +45:53:25.31 & 5.08$\pm$0.03 & 1.8$\pm$0.1 & &  & & & 21 $\pm$ 5 & -26.00$^{+0.02}_{-0.04}$& 7.2 $^{+ 1.9 } _{- 2.2 }$ & 43.05 $^{+ 0.53 }_{- 0.54 }$ \\
ILTJ1104+6631  & 11:04:55.18 +66:31:18.62 & 5.03$\pm$0.06 & 0.5$\pm$0.2 & &  & & & 5 $\pm$ 3& -26.72$^{+0.02}_{-0.01}$ & 21 $^{+ 1 } _{- 4 }$ & 44.06 $^{+ 0.11 }_{- 0.12 }$ \\
ILTJ1133+4814  & 11:33:50.43 +48:14:31.22 & 6.25$\pm$0.02 & 4.6$\pm$0.3 & 3.23$\pm$0.14 & 2.54$\pm$0.29 & $-0.16\pm$0.04 & $-0.32\pm$0.16 & 360 $\pm$ 100 & -25.00$^{+0.20}_{-0.40}$& 38 $^{+ 3 } _{- 2 }$ & 43.79 $^{+ 0.04 }_{- 0.03 }$ \\ 
ILTJ1330+6224  & 13:30:28.26 +62:24:12.28 & 5.43$\pm$0.02 & 0.8$\pm$0.2 & &  & & & 16 $\pm$ 6 & -26.25$^{+0.04}_{-0.08}$& 40 $^{+ 2} _{- 4 }$ & 44.13 $^{+ 0.2 }_{- 0.21 }$ \\
ILTJ1334+4750  & 13:34:22.64 +47:50:33.50 & 4.99$\pm$0.03 & 5.5$\pm$0.1 & 0.54$\pm$0.16$^{b}$ & & $-1.02\pm$0.13 & & 10 $\pm$ 3 & -27.02$^{+0.01}_{-0.01}$ & 1.9 $^{+ 0.1 } _{- 0.2 }$ & 43.10 $^{+ 0.43 }_{- 0.43 }$ \\
ILTJ1350+3748  & 13:50:23.60 +37:48:35.72 & 5.72$\pm$0.03 & 0.5$\pm$0.2 & &  & & & 17 $\pm$ 7 &  -26.00$^{+0.04}_{-0.04}$& 4.4 $^{+ 0.5 } _{- 0.5 }$ & 42.95 $^{+ 0.22 }_{- 0.22 }$ \\
ILTJ1401+4542 &14:01:20.86 +45:42:53.46 & 5.45$\pm$0.04 & 0.4$\pm$0.1 & &  & & & 10 $\pm$ 3 & -25.50$^{+0.03}_{-0.03}$ & 14 $^{+ 1 } _{- 1 }$ & 43.24 $^{+ 0.15 }_{- 0.15 }$ \\
ILTJ1419+5718  & 14:19:56.58 +57:18:40.41 & 5.35$\pm$0.04 & 0.5$\pm$0.1 & &  & & & 21 $\pm$ 9 & -25.77$^{+0.08}_{-0.04}$& 6.4 $^{+ 9.7 } _{- 2.1 }$ & 43.15 $^{+ 0.74 }_{- 0.59 }$ \\
ILTJ1523+2935 & 15:23:30.67 +29:35:39.67 & 5.74$\pm$0.03 & 1.3$\pm$0.2 & & 0.39$\pm$0.10$^{b}$ & $-0.38\pm$0.10$^{c}$ & & 30 $\pm$ 12 & -26.32$^{+0.02}_{-0.02}$ & 36 $^{+ 2 } _{- 1 }$ & 44.01 $^{+ 0.23 }_{- 0.23 }$  \\
ILTJ1601+3102  & 16:01:49.45 +31:02:07.25 & 4.90$\pm$0.02 & 69$\pm$0.7 & 3.74$\pm$0.13 & 1.68$\pm$0.29 & $-1.28\pm$0.02 & -1.05$\pm$0.23 & 1000 $\pm$ 400 & -24.75$^{+0.31}_{-0.21}$&  19 $^{+ 2 } _{- 3 }$ & 43.25 $^{+ 0.31 }_{- 0.31 }$ \\ 
ILTJ1650+5457  & 16:50:51.75 +54:57:01.38 & 6.06$\pm$0.02 & 0.6$\pm$0.3 & &  & & & 10 $\pm$ 6 & -26.56$^{+0.04}_{-0.08}$& 144 $^{+ 14 } _{- 12 }$ & 44.47 $^{+ 0.27 }_{- 0.27 }$ \\ 
ILTJ2201+2338  & 22:01:07.61 +23:38:37.95 & 5.83$\pm$0.02 & 3.6$\pm$0.4 &  & 0.42$\pm$0.10$^{b}$ & $-0.71\pm$0.08$^{c}$ & & 37 $\pm$ 11 & -26.24$^{+0.06}_{-0.11}$&  35 $^{+ 1 } _{- 1 }$ & 44.07 $^{+ 0.01 }_{- 0.01 }$ \\
ILTJ2327+2454  & 23:27:26.89 +24:54:10.93 & 5.30$\pm$0.03 & 1.4$\pm$0.2 & &  & & & 10 $\pm$ 3 & -26.07$^{+0.02}_{-0.05}$& 19 $^{+ 1 } _{- 1 }$ & 43.87 $^{+ 0.14 }_{- 0.14 }$\\
ILTJ2336+1842  & 23:36:24.69 +18:42:48.71 & 6.60$\pm$0.04 & 1.4$\pm$0.4 & &  & & & 59 $\pm$ 26 & -24.32$^{+0.13}_{-1.44}$& 33 $^{+ 6 } _{- 3 }$ & 43.92 $^{+ 0.2 }_{- 0.19 }$\\
\hline \hline 
\vspace{0.5pt}
\end{tabular}}
\raggedright \small{\newline  \textbf{Notes.} $^a$ A redshift of $z=5.24\pm0.01$ is obtained from fitting the O\textsc{i} and Si \textsc{iv} emission lines and a redshift of $5.34\pm0.02$ is obtained from template fitting. $^b$ These sources are not included in the FIRST or VLASS catalogue, but are marginally detected (3-6$\sigma$) in the images. The values given here are the peak fluxes including a 15\% correction for the VLASS measurements. $^c$ These spectral indices are calculated from the LOFAR and VLASS survey measurements ($\alpha_{144\text{MHz}}^{3\text{GHz}}$). The full table is available in electronic form at the CDS via anonymous ftp to cdsarc.u-strasbg.fr (130.79.128.5) or via \url{http://cdsweb.u-strasbg.fr/cgi-bin/qcat?J/A+A/} or \url{https://doi.org/10.5281/zenodo.7142693}. \\}
\end{table*}

\subsection{Low-frequency radio properties}
\label{subsec:low_freq_prop}

\begin{figure}
\centering
   \includegraphics[width=\columnwidth, trim={0.0cm 0cm 0cm 0.0cm}, clip]{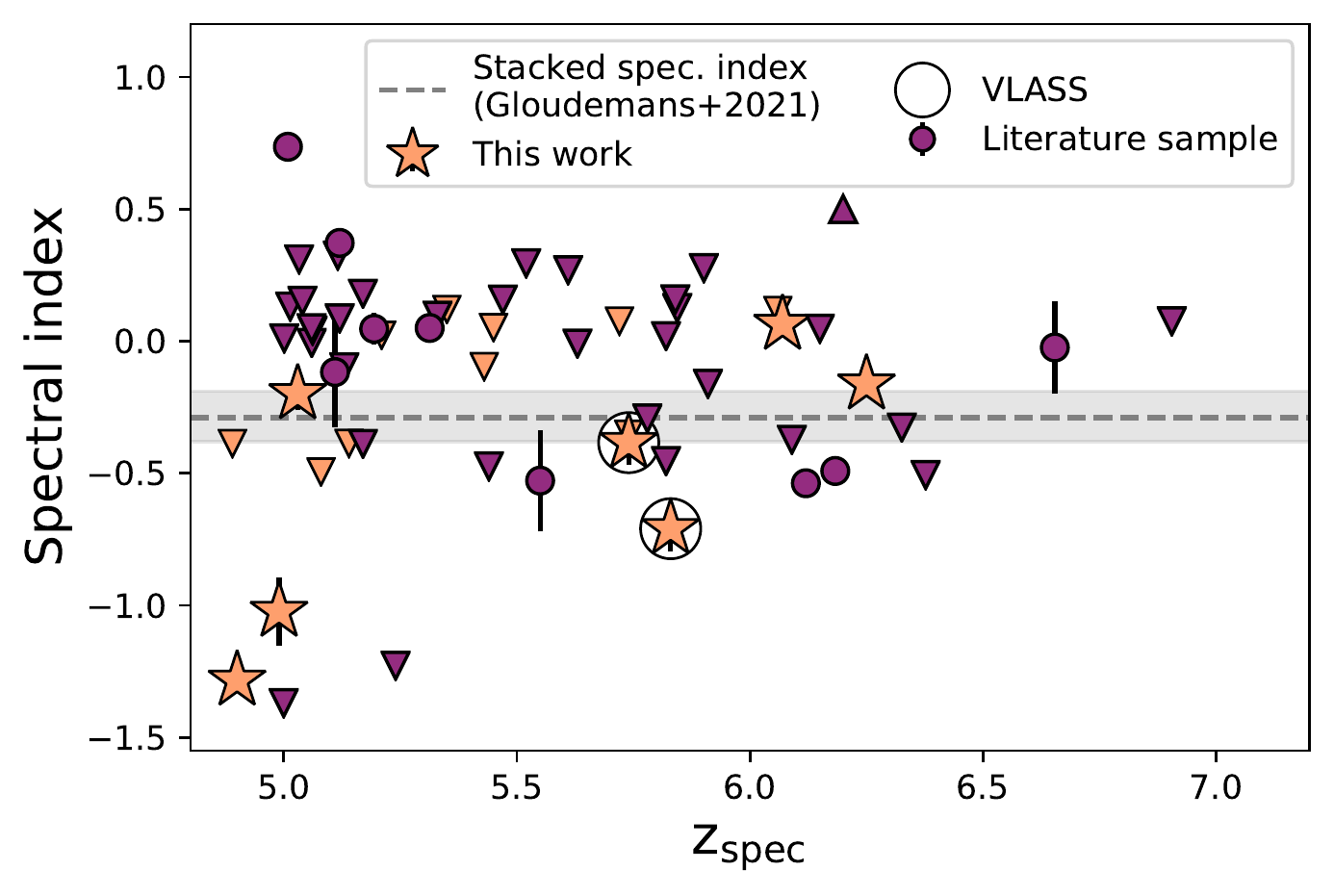}
     \caption{Radio spectral index between 144 MHz (LoTSS-DR2) and 1.4 GHz (VLA FIRST) for the discovered quasars in the FIRST and VLASS survey area. Upper limits (indicated by downward-pointing triangles) are given for the quasars non-detected in FIRST. The dashed grey line shows the stacked spectral index obtained by \cite{Gloudemans2021A&A...656A.137G} using a sample of 115 known quasars at $z>5$. ILTJ2201+2338 is outside of the FIRST footprint, but is detected in VLASS at 2-4 GHz, which enables a spectral index measurement. ILTJ1523+2935 is only marginally detected in FIRST (S/N$\sim$2), but is more securely detected in VLASS (S/N$=3.4$).}
     \label{fig:radio_spectral_index}
\end{figure}

\begin{figure}
\centering
   \includegraphics[width=\columnwidth, trim={0.0cm 0cm 0cm 0.0cm}, clip]{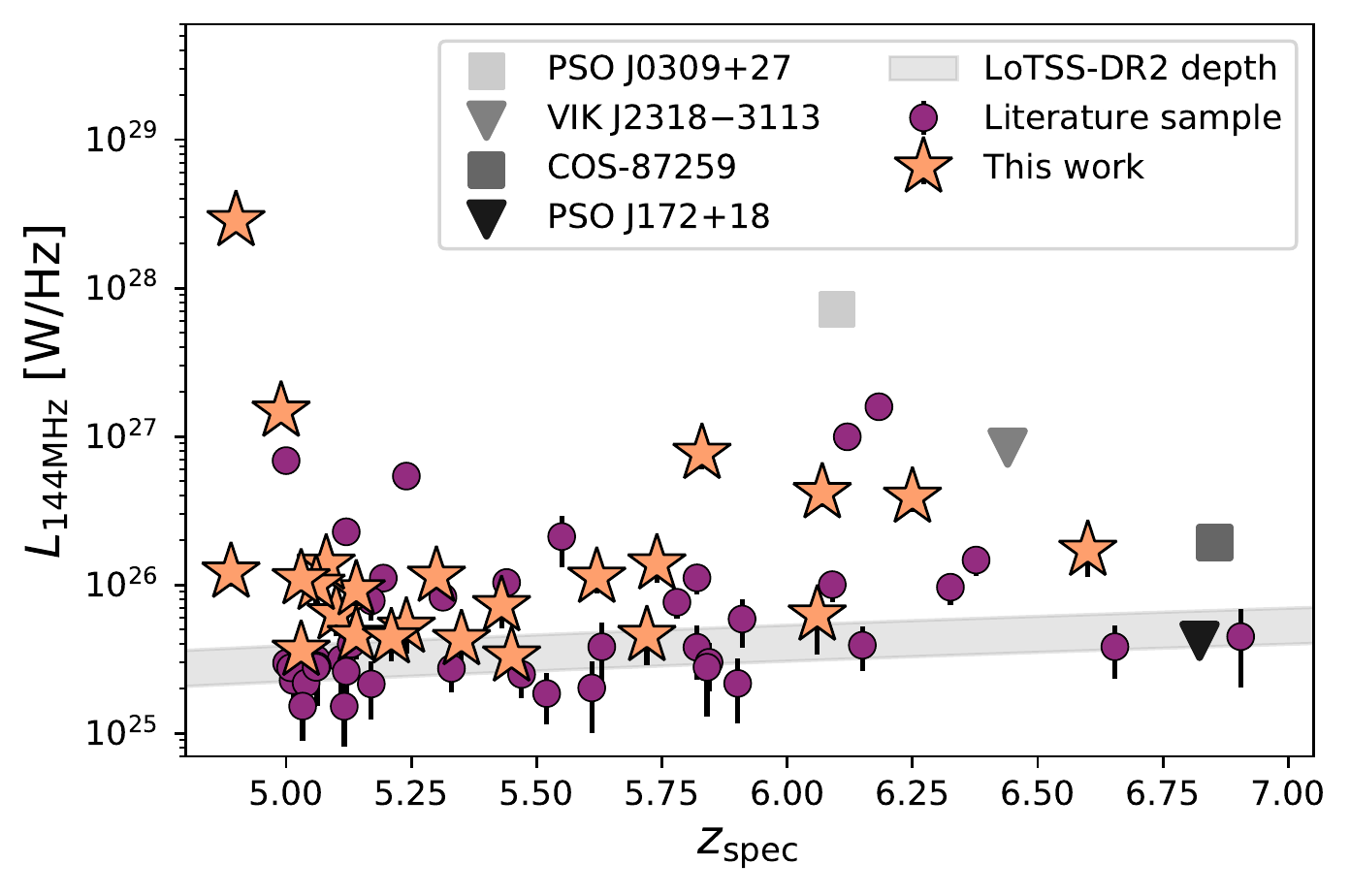}
     \caption{Radio luminosities at 144 MHz of newly discovered quasars in this work compared to the literature sample of known high-$z$ quasars detected in LoTSS-DR2. Other known $z>6$ radio-loud quasars PSO J172+18 \citep{Banados2021ApJ...909...80B} and VIK J2318-3113 \citep{Ighina2021A&A...647L..11I} remain non-detected and not covered by LOFAR, respectively. PSO J0309+27 \citep{Belladitta2020A&A...635L...7B} and COS-87259 \citep{Endsley2022arXiv220600018E} are detected by LOFAR outside of the LoTSS-DR2 area.}
     \label{fig:radio_lum} 
\end{figure}

\begin{figure*}
    \centering
    \begin{minipage}{0.48\linewidth}
        \centering
        \includegraphics[width=\columnwidth, trim={0.0cm 0cm 0cm 0.0cm}, clip]{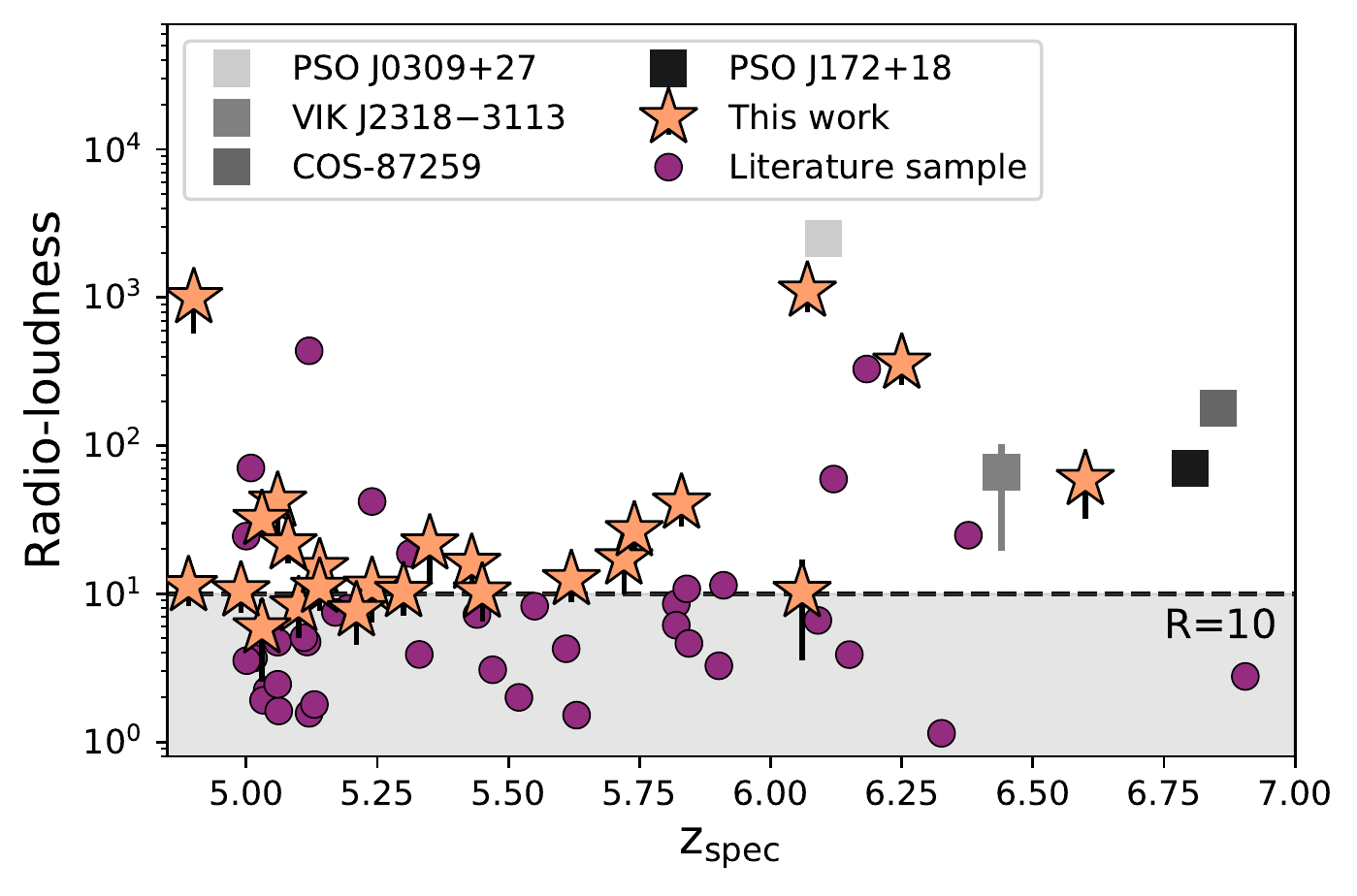}
    \end{minipage}
    \begin{minipage}{0.48\linewidth}
        \centering
        \includegraphics[width=1.0\linewidth, trim={0cm 0cm 0.0cm 0.0cm}, ]{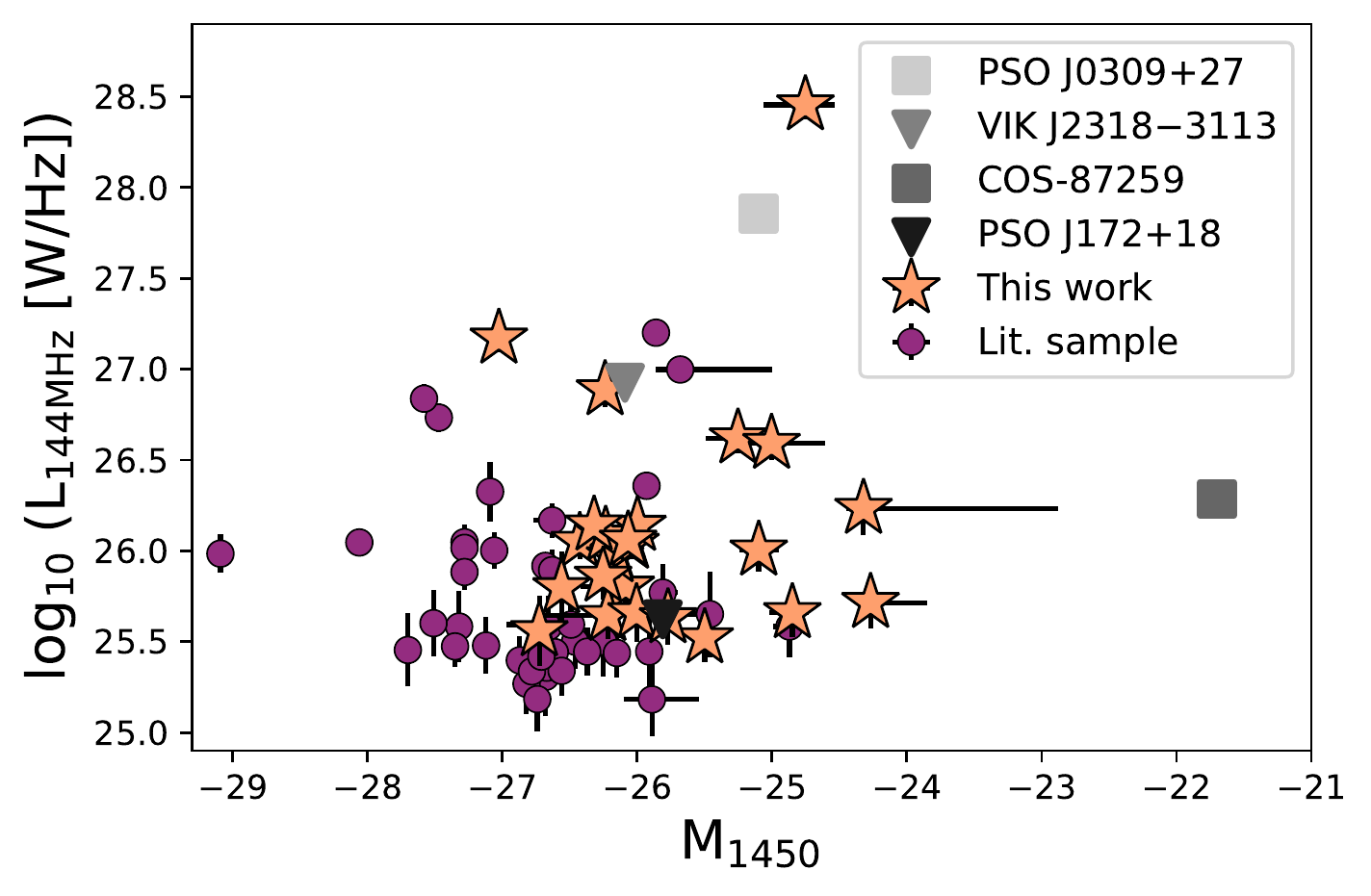}
    \end{minipage}
    \vspace{-10pt}
    \caption{\label{fig:radio_loudness}\small{{\textbf{Left:} Radio-loudness ($R=f_{5\text{GHz}}/f_{4400\text{\AA}}$) as a function of spectroscopic redshift for the discovered quasars in this work and the literature quasar sample. Our quasars span a broad range in radio-loudness. Traditionally, quasars are defined as radio-loud if $R>10$, which is satisfied by 21/24 of our newly discovered quasars. \textbf{Right:} UV rest-frame absolute magnitude at 1450 \r{A} versus radio luminosity at 144 MHz. The new quasar sample probes fainter UV magnitudes and higher average radio luminosities than previously known quasar samples.}}}
\end{figure*}

To measure the radio luminosities and radio-loudness of the new quasar sample from the observed LOFAR fluxes, we need to determine their radio spectral indices (using a simple power-law of $S_{\nu} = \nu^{\alpha}$). Previous work of \cite{Gloudemans2021A&A...656A.137G} has shown that the median spectral index of the known quasar population between 144 MHz and 1.4 GHz is approximately $-0.3\pm0.1$. To obtain spectral indices for our newly discov- ered quasars we cross-match our sources (within 2$\arcsec$) with the VLA FIRST survey at 1.4 GHz \citep{Becker1994ASPC...61..165B}, which yields detections for 4 quasars. Additionally, to be able to also constrain the spectral index on marginally detected sources we perform forced photometry by taking the peak flux, which results in 1 additional $\sim$3$\sigma$ measurement of ILTJ1334+4750.  

The spectral indices derived from these detections are shown in Fig.~\ref{fig:radio_spectral_index}, with 3$\sigma$ upper limits derived for the non-detected quasars ($<3\sigma$) in FIRST. We could not obtain a measurement for nine out of 24 quasars, which were outside of the FIRST survey footprint. We repeat this procedure using the Very Large Array Sky Survey (VLASS; \citealt{Condon1998AJ....115.1693C}) at 2-4 GHz, which yields 3 detected sources in the VLASS catalogue and three marginally detected sources from measuring the peak flux. Since the VLASS images are known to systematically underestimate the flux densities by $\approx$15\%, we have applied this correction to our peak fluxes. These six detected quasars in VLASS resulted in additional spectral index measurements for ILTJ1523+2935 and ILTJ2201+2338 (see Fig. ~\ref{fig:radio_spectral_index}). All measured fluxes and spectral indices are given in Tab.~\ref{tab:quasar_properties}. From this Fig.~\ref{fig:radio_spectral_index} it is apparent there is a large scatter in spectral index around the median value of $-0.3\pm$0.1, which is determined in \cite{Gloudemans2021A&A...656A.137G} by median stacking LOFAR and FIRST images of 93 known quasars at $z>5$. Therefore, in this work we assume $\alpha=-0.3\pm0.1$ if there is no FIRST or VLASS detection. Finally, we note some of these quasars could also be peaked-spectrum sources with a turnover in the radio spectrum (see e.g. review by \citealt{Urry1995PASP..107..803U}). However, the spectral indices derived from VLASS for the other four quasars agree very well with the FIRST spectral indices (with differences of 0.04-0.15), indicating no significant spectral curvature for these sources at GHz frequencies. For the other quasars more radio observations are necessary to constrain the radio spectrum shape.

The radio luminosities of the newly discovered quasars as a function of redshift are shown in Fig.~\ref{fig:radio_lum}. Our sample probes on average higher radio luminosities than the previously known quasars, but most of the brightest radio sources in our candidate sample were identified as low-$z$ contaminants. However, the quasar ILTJ1601+3102 at $z=4.9$ has an exceptionally high radio flux of 69 mJy. Besides the known quasar sample in LoTSS-DR2 from \cite{Gloudemans2021A&A...656A.137G}, we also add other recently published radio-loud quasars: PSO J172+18 \citep{Banados2021ApJ...909...80B}, PSO J0309+27 \citep{Belladitta2020A&A...635L...7B}, VIK J2318−3113 \citep{Ighina2021A&A...647L..11I}, and COS-87259 \citep{Endsley2022arXiv220600018E}. PSO J0309+27 has been detected by LOFAR outside of the current LoTSS-DR2 area with a radio flux of 52 mJy at 144 MHz (Spingola et al. in prep) and has been classified as blazar by \cite{Belladitta2020A&A...635L...7B}. The highest redshift radio-loud quasar COS-87259 has also been detected by LOFAR outside of the LoTSS-DR2 footprint with a 144 MHz flux of 0.48$\pm$0.18 mJy \citep{Endsley2022MNRAS.512.4248E}. The other very high-$z$ redshift radio-loud quasar PSO J172+18 at $z=6.8$ remains non-detected by LOFAR with an RMS noise limit of $\sim$0.3 mJy. This non-detection by LOFAR puts a new constraint on the radio spectral index of $\alpha > 0.2$ using the VLA-L band measurement at 1.52 GHz of $0.51\pm0.02$ mJy (see \citealt{Banados2021ApJ...909...80B}), since it was previously only non-detected by TGSS with a flux limit of 8.5 mJy. The spectral index between the VLA-L and VLA-S band was measured to be negative with $\alpha^{L}_{S} = -1.31 \pm 0.08$. This quasar therefore shows a turnover in the spectrum. The quasar VIK J2318$-$3113 was the previous highest redshift radio-loud quasar, but this region is not observable by LOFAR due to the low declination and therefore a FIRST flux upper limit is given here.

The radio-loudness of quasars is traditionally defined by the ratio of the radio (5 GHz) and optical (4400 \r{A}) flux densities (radio-loudness criterion; $R=f_{5\text{GHz}}/f_{4400\text{\AA}} > 10$; e.g. \citealt{Kellermann1989AJ.....98.1195K}). Therefore, to compare the radio-loudness of our new quasars with the known quasar population we estimate the radio flux at 5 GHz rest-frame by extrapolating from the observed 144 MHz flux ($\sim 850-1100$ MHz in rest-frame) assuming our determined radio spectral indices from Fig.~\ref{fig:radio_spectral_index}. The resulting radio-loudness ($R$) compared to known quasars is shown in the left panel of Fig.~\ref{fig:radio_loudness}. The radio-loudness of our quasars is in the range of 6-1080. Traditionally, quasars are defined as radio-loud if $R>10$, which is satisfied by 21 out of the 24 quasars with 23 out of the 24 quasars having $R>10$ within the error margin. The blazar PSO J0309+27 at $z=6.1$ is still the most radio-loud quasar-like source at $z>6$. We note that this result is highly dependent on the assumed radio spectral indices, which could not be determined for the full sample. For our sample of quasars a change of 0.1 in the spectral index gives rise to a difference of $\sim$16-19\% in the radio-loudness. These errors are taken into account in Fig.~\ref{fig:radio_loudness}. Follow-up radio observations at GHz frequencies are necessary to constrain the spectral indices (and potential curvature) and accurately determine the radio-loudness. This is also the case for the LOFAR detected known quasar sample, for which \cite{Gloudemans2021A&A...656A.137G} assumed a spectral index of $-0.3\pm0.1$ for all non-detected quasars in the FIRST survey.

As shown in the right panel of Fig.~\ref{fig:radio_loudness}, our quasars probe a new parameter space of fainter rest-frame UV magnitudes and on average higher radio luminosities (at 144 MHz) than previously known LOFAR detected high-$z$ quasars. To compare, the new quasars have a median L$_{144\text{MHz}}$ and M$_{1450}$ of 10$^{26.02}$ W Hz$^{-1}$ and -26.06, respectively, whereas the known quasars detected by LOFAR have a median L$_{144\text{MHz}}$ and M$_{1450}$ of 10$^{25.58}$ W Hz$^{-1}$ and -26.68 respectively. Our quasars are still not radio bright enough ($>10$ mJy) at high enough redshift ($z>6$) for current facilities such as LOFAR to be able to measure the 21-cm absorption line. However, these measurements might be possible with similar quasars in the Southern Hemisphere with the future Square Kilometer Array (see \citealt{ciardi2015aska.confE...6C}). We note that populations of fainter quasars have been found at $z > 5.7$ in optical surveys such as the Subaru High-$z$ Exploration of Low-Luminosity Quasars (SHELLQs; e.g. \citealt{Matsuoka2016ApJ...828...26M, Matsuoka2022ApJS..259...18M}), but these generally remain non-detected by LOFAR \citep{Gloudemans2021A&A...648A...7G}.  

Other recent studies in the radio regime have found that about 10\% of the high-$z$ quasars are radio-loud using the classical definition of the ratio of the radio and optical fluxes $f_{5\text{GHz}}$/$f_{4400\text{\AA}} > 10$ \citep{Banados2015ApJ...804..118B, Gloudemans2021A&A...656A.137G, Liu2021ApJ...908..124L}. However, in the current study we have selected quasar candidates that are relatively radio bright and therefore our fraction of radio-loud quasars is significantly higher. There is still ongoing discussion on whether the classes of radio-loud and -quiet quasars are probing different populations, since different studies find both evidence for bimodality (e.g. \citealt{Ivezic2002AJ....124.2364I, White2007ApJ...654...99W, Zamfir2008MNRAS.387..856Z, Beaklini2020MNRAS.497.1463B}) and against bimodality (e.g. \citealt{Cirasuolo2003MNRAS.346..447C, Balokovic2012ApJ...759...30B}) of the radio luminosity distribution of quasars. Recently, \cite{gurkan2019A&A...622A..11G} showed that this bimodality is not favoured by the data using a large sample of quasars ($\sim$50,000) covered by the 150 MHz LOFAR surveys. Also, recent work of \cite{Macfarlane2021} shows that the radio luminosity distribution can be modelled using a combination of radio emission from jets (with a wide distribution of radio powers) and star formation with a smooth transition between the radio-loud and -quiet regime and without the need for bimodality.

\subsection{Quasar colours}
\label{subsec:colour_prop}

\begin{figure*}
\centering
   \includegraphics[width=\textwidth, trim={0.0cm 0cm 0cm 0.0cm}, clip]{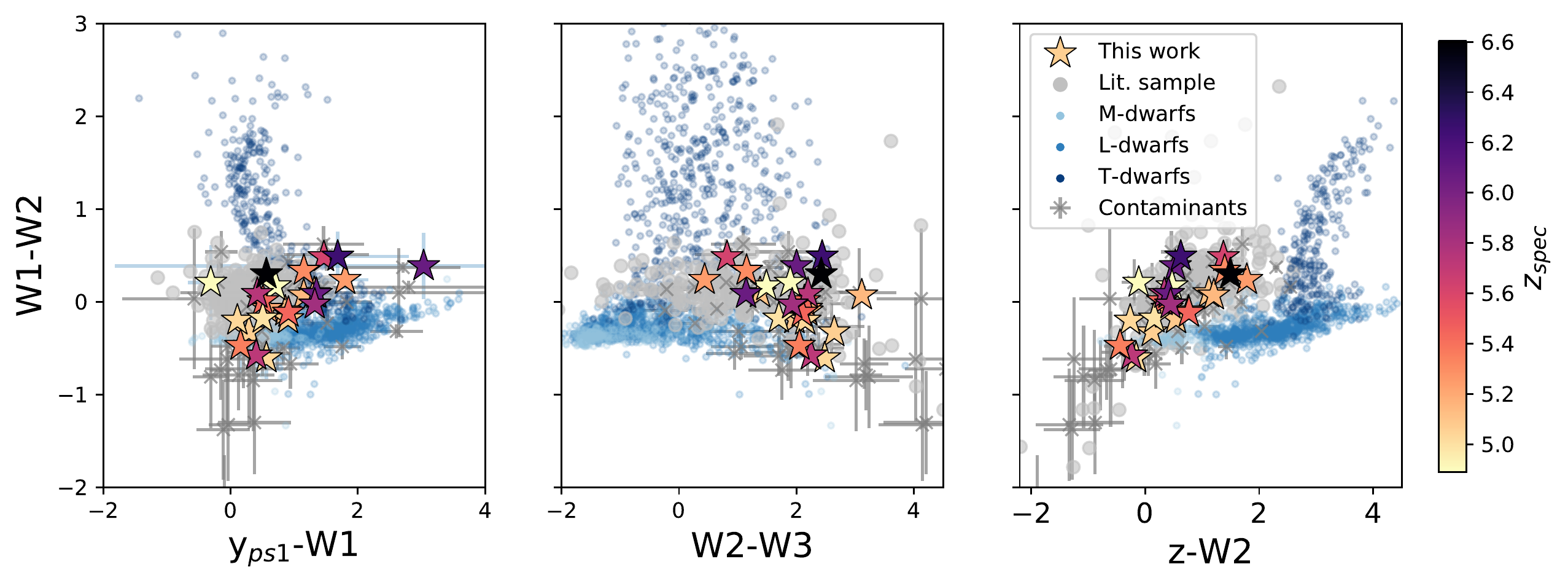}
     \caption[]{Near-infrared colours of the discovered quasars in this work and contaminants compared to the literature sample of quasars and stellar dwarfs\protect\footnotemark (for which the PS-1 $z$-band magnitude is used in the right panel). Despite a looser colour selection, the new quasars exhibit colours consistent with the known quasar population.}
     \label{fig:wise_colors}
\end{figure*}

\footnotetext{\url{https://doi.org/10.5281/zenodo.4169085}}

With this campaign we have selected quasars in a broad redshift range ($4.9 \lesssim z \lesssim 6.6$) and colour space (see Section~\ref{subsec:colour_selection}), whereas previous high-$z$ quasar searches have focused on for example $4.7 < z < 5.4$ (e.g. \citealt{Wang2016ApJ...819...24W}), $z \gtrsim 5.6$ (e.g. \citealt{Banados2016ApJS..227...11B, Jiang2016ApJ...833..222J, Wang2019ApJ...884...30W}), or quasars in the redshift gap between $z \sim $5.3-5.7 (e.g. \citealt{YangJ2017, Yang2019ApJ...871..199Y}). Despite selecting our quasar candidates using fewer and less conservative colour restrictions, our overall success rate is 34\% and we have discovered 7 new quasars in the redshift gap. In other high-$z$ quasar searches the success rate ranges between $\sim$5-64\% depending on the selection method and redshift range (see e.g. \citealt{ Wang2016ApJ...819...24W,YangJ2017,Matsuoka2019ApJ...883..183M, Wang2019ApJ...884...30W, Nanni2022MNRAS.515.3224N,Wagenveld2022A&A...660A..22W}), however this number is not always published. Our success rate can be attributed to the radio detections drastically decreasing the number of contaminants in our sample and this work therefore demonstrates that a relatively clean high-$z$ quasar candidate sample can be created using radio detections without requiring the full set of optical and NIR filters normally employed at these redshifts. We note the WISE photometry has also contributed significantly in constraining the photometric redshifts in this work. 

The optical and near-infrared colours of the new quasars do not significantly deviate from the known quasar population, as is apparent in Fig.~\ref{fig:selection_colors} and \ref{fig:wise_colors}. All candidates that were spectroscopically observed outside of the traditional colour space have been identified as low-$z$ contaminants, which are discussed in Appendix \ref{sec:appendix_contaminants}.

\subsection{Rest-frame UV spectral properties}
\label{subsec:optical_prop}

\begin{figure*}
    \centering
    \begin{minipage}{0.47\linewidth}
        \centering
        \includegraphics[width=\columnwidth, trim={0.0cm 0cm 0cm 0.0cm}, clip]{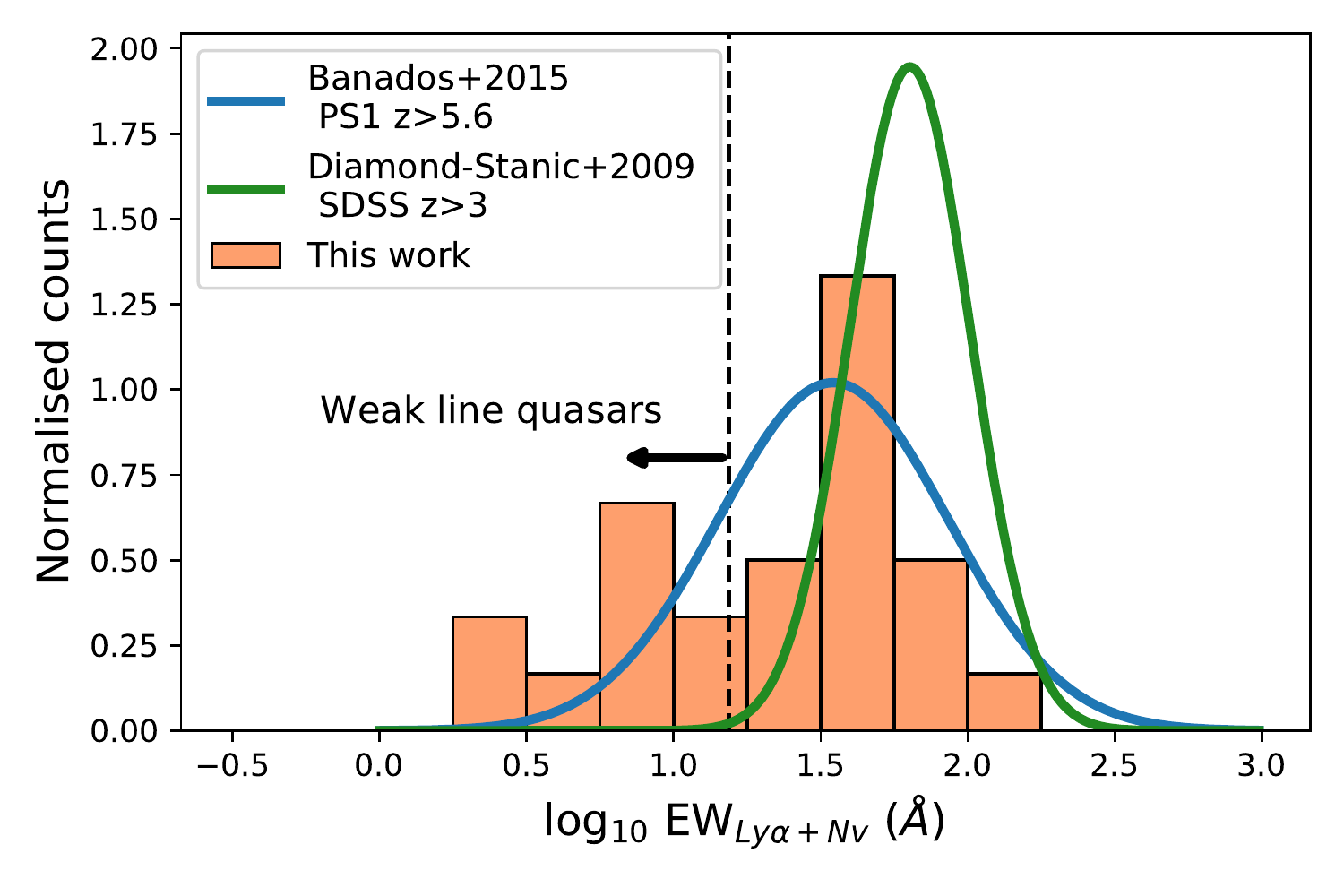}
    \end{minipage}
    \begin{minipage}{0.49\linewidth}
        \centering
        \includegraphics[width=1.0\linewidth, trim={0cm 0cm 0.0cm 0.0cm}, ]{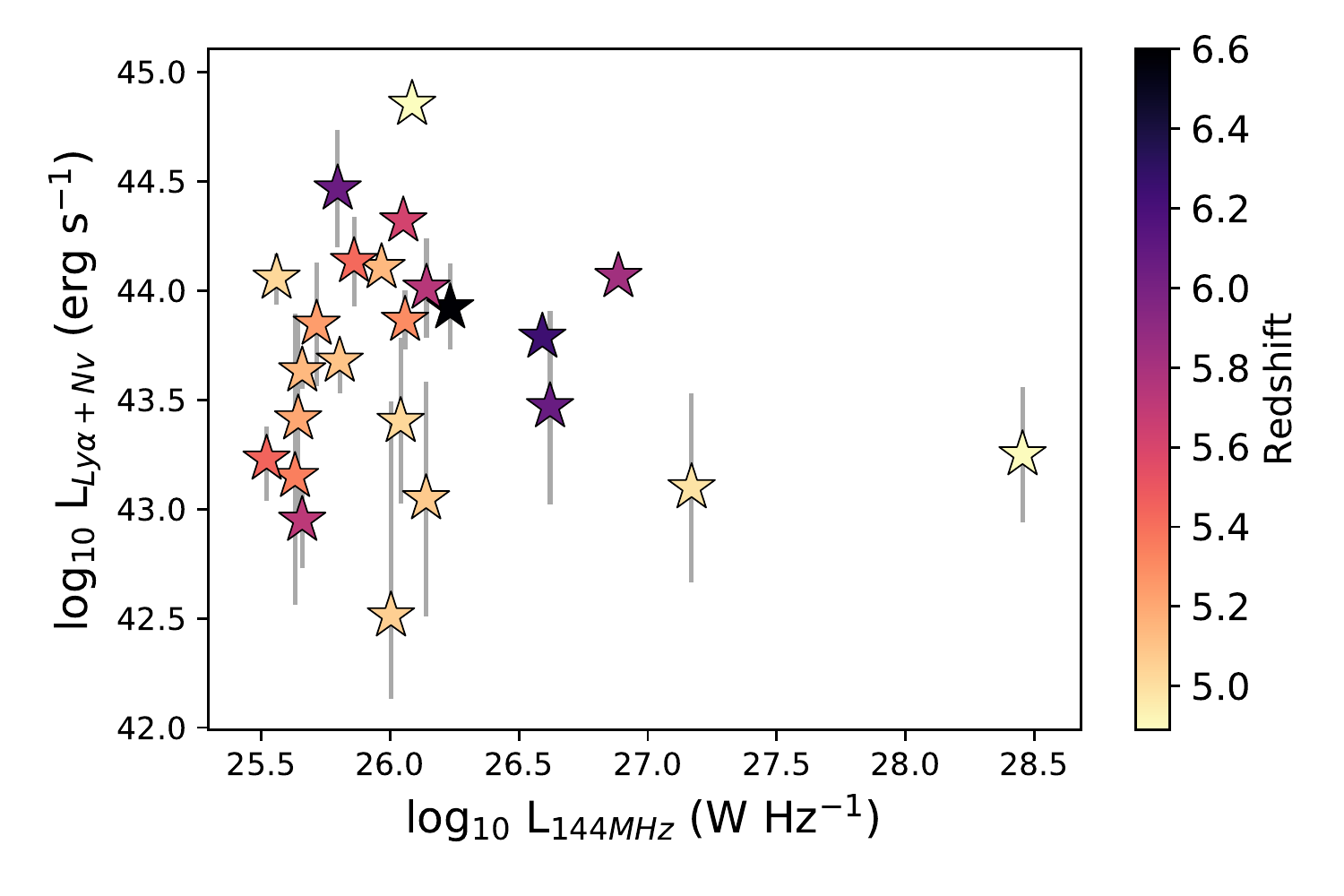}
    \end{minipage}
    \vspace{-10pt}
    \caption{\label{fig:Lya_EW}\small{{Derived Ly$\alpha$+N\textsc{v} emission line properties of the discovered quasars \textbf{Left:} Normalised Ly$\alpha$+N\textsc{v} EW distribution of the discovered quasars in this work compared to the distribution of PS1 discovered quasars at $z>5.6$ by \cite{Banados2016ApJS..227...11B} and SDSS discovered quasars at $z>3$ by \cite{Diamond-Stanic2009ApJ...699..782D}. Our EW distribution is consistent with the optically selected quasars sample of \cite{Banados2016ApJS..227...11B}. \textbf{Right:} Ly$\alpha$+N\textsc{v} luminosity versus the radio luminosity at 144 MHz for the quasars in this work, showing little to no correlation (see Section~\ref{subsec:optical_prop}).}}}
\end{figure*}

In this section, we compare the Ly$\alpha$+N\textsc{v} line strengths of our RL quasars to known RQ quasars to investigate whether they may probe different populations. We compare our EW distribution to the distribution obtained by \cite{Banados2016ApJS..227...11B} for PS1 selected quasars at $z>5.6$ in Fig.~\ref{fig:Lya_EW}. A Kolmogorov-Smirnov (KS) test yields a p-value of 0.10 indicating no statistically significant deviation between our EW distribution and the PS1 selected distribution from \cite{Banados2016ApJS..227...11B}. As stated in \cite{Banados2016ApJS..227...11B}, the EW distribution determined for SDSS selected quasars at $z>3$ from \cite{Diamond-Stanic2009ApJ...699..782D} is likely systematically higher due to lower IGM absorption of Ly$\alpha$ at lower redshift. From our sample, 9 out of 24 quasars (38\%) have EW values below the defined weak line quasar boundary of 15.4 \r{A} (see \citealt{Diamond-Stanic2009ApJ...699..782D}), which is almost 3 times higher than the fraction of weak line quasars of 14\% (16/117) reported from the PS1 quasar sample of \cite{Banados2016ApJS..227...11B}. This could indicate a potential difference in observed Ly$\alpha$ emission between radio-quiet and -loud high-$z$ quasars, but these are still low number statistics. However, the consistency of the overall shape the EW distribution of our radio-loud quasar sample with the optically selected quasars from \cite{Banados2016ApJS..227...11B} potentially suggests there is no strong radio-loudness dichotomy between the radio-loud and radio-quiet quasar population in the sense that the optical quasar properties are similar. 

A previous study of high-$z$ radio galaxies (HzRGs) by \cite{Jarvis2001MNRAS.326.1563J} found a strong positive correlation between $L_{\text{Ly}\alpha}$ and $L_{151}$ for their sample of 35 HzRGs at $z>1.75$, however \cite{saxena2019MNRAS.489.5053S} only found a weak positive correlation, consistent with no correlation, for their sample consisting of 10 new HzRGs at $z>2$ and literature HzRGs. To investigate if such a correlation exists for our quasar sample, we plot the radio luminosity against the Ly$\alpha$ luminosity in the right panel of Fig.~\ref{fig:Lya_EW}. We find a weak negative correlation between the Ly$\alpha$ and radio luminosity with a Pearson correlation coefficient of $r=-0.12$. We find a very weak negative evolution of Ly$\alpha$ luminosity with redshift ($r=-0.05$), but a slightly stronger correlation of radio luminosity with redshift ($r=-0.24$). However, the results from this quasar sample cannot be directly compared to the radio galaxies, since the rest-frame UV and radio emission we are probing are driven by accretion that is varying on shorter timescales than radio galaxies, which generally have more extended radio-lobes (see e.g. \citealt{Miley2008A&ARv..15...67M, saxena2019MNRAS.489.5053S, Nyland2020ApJ...905...74N}). 

Although we do not see any strong correlation between Ly$\alpha$ and radio luminosity from our new sample of faint quasars, we note that any interpretation is limited by our relatively small sample size. In addition, interpreting the emission line properties of Ly$\alpha$ at high redshift is complicated, since resonant scattering becomes increasingly important, and therefore the line profile depends on both the intrinsic emission and the environment of the quasar or HzRG (e.g. \citealt{Ojik1997A&A...317..358V}). Future NIR spectroscopy (targeting broad emission lines such as C\,\textsc{iv} and Mg\,\textsc{ii}) will enable the measurement of SMBH masses and Eddington ratios of these sources (see e.g. \citealt{Onoue2019ApJ...880...77O}) to constrain the timescales of SMBH growth and systemic redshifts to characterise their Ly$\alpha$ damping wings and outflows. 

\subsection{Red quasars}
\label{subsec:redquasars}

\begin{figure}
\centering
   \includegraphics[width=\columnwidth, trim={0.0cm 0cm 0cm 0.0cm}, clip]{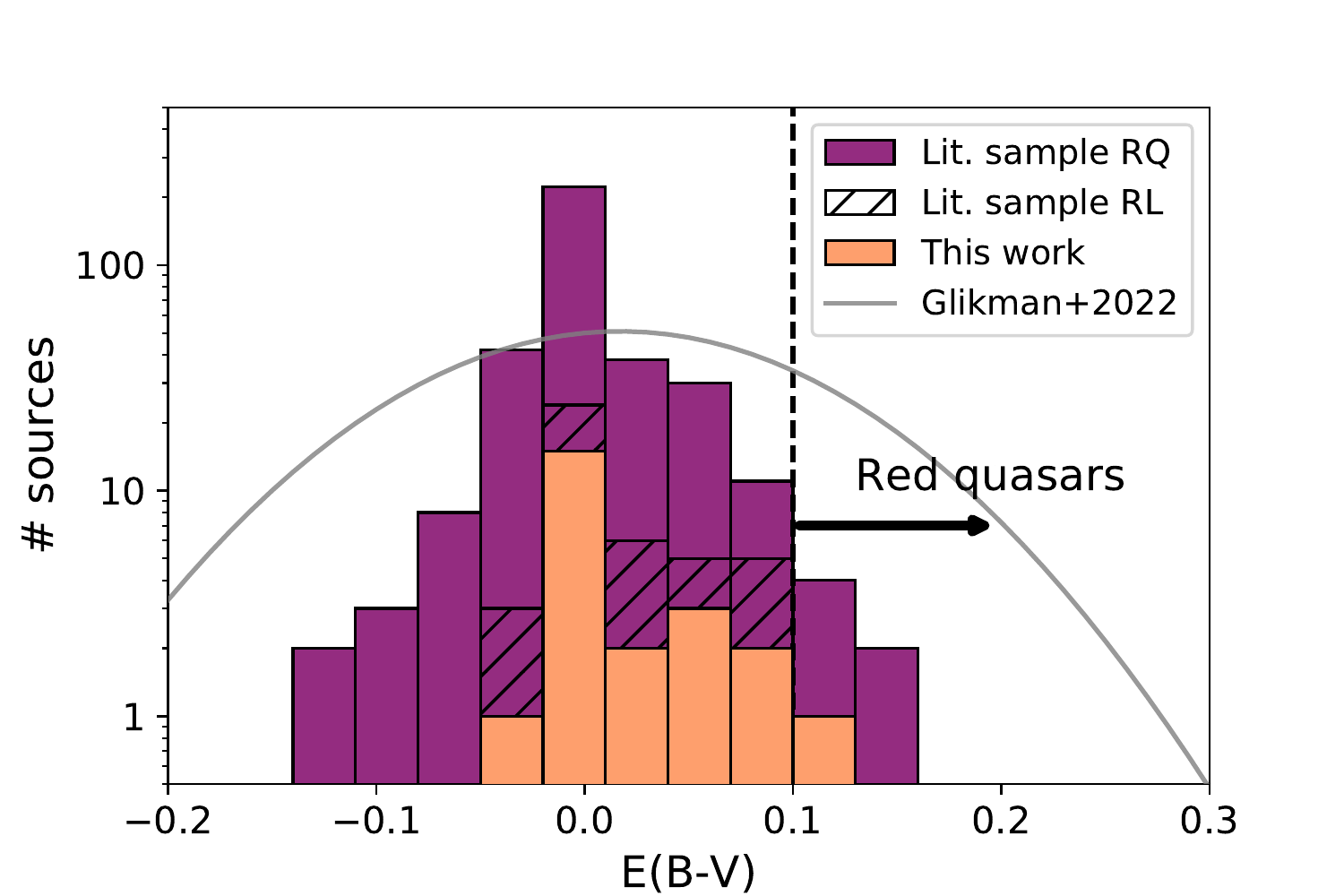}
     \caption{Distribution of $E(B-V)$ for the discovered quasars in this work and the literature sample determined by fitting the composite quasars spectrum of \cite{Selsing2016A&A...585A..87S} together with the SMC extinction law \citep{Pei1992ApJ...395..130P}. The radio-loud quasars from the literature sample are stacked on top of quasars from this work. We do not find a significant deviation between the E(B-V) distribution of the RL and RQ quasar sample. For comparison we also plot the $E(B-V)$ Gaussian distribution found by \cite{Glikman2022ApJ...934..119G} of SDSS QSOs at $0.1 < z \lesssim 3$ using a similar template fitting method. }
     \label{fig:e_b_v_plot}
\end{figure}

Following \cite{Glikman2012ApJ...757...51G}, high-$z$ quasars can be classified as red when the dust reddening parameter $E(B-V) > 0.1$. The dust reddened red quasar population has been shown to have an enhanced fraction of radio-loud quasars (e.g. \citealt{Fawcett2020MNRAS.494.4802F, Rosario2020MNRAS.494.3061R, Glikman2022ApJ...934..119G}). Therefore, to investigate whether our radio selection combined with loose optical colour criteria has resulted in an increased fraction of red quasars, we examine the dust reddening properties of our sample. Previous observational studies using the Atacama Large Millimeter Array (ALMA) have found abundant dust in quasar host galaxies (see e.g. \citealt{Wang2013ApJ...773...44W,Venemans2016ApJ...816...37V, Izumi2019PASJ...71..111I}), however the fraction of extinguished quasar radiation caused by dust and whether this represents an evolutionary stage in a quasar's lifetime is still unknown. Furthermore, the FIR and millimetre emission of radio-loud quasars is likely not only caused by cold dust but also by synchrotron emission \citep{Rojas2021ApJ...920..150R,Khusanova2022A&A...664A..39K}. 

To determine the $E(B-V)$ for our quasars, we used a template fitting method on the available optical and (N)IR photometry. We fitted the composite quasar spectrum of \cite{Selsing2016A&A...585A..87S} to the photometric data points redward of the Lyman break and apply both the SMC extinction law of \cite{Pei1992ApJ...395..130P} with $R_V = 2.93$ and the starburst galaxies extinction curve of \cite{Calzetti2000ApJ...533..682C} with $R_V = 3.1$ for different values of $E(B-V)$. The SMC dust reddening law is assumed to be similar to high-$z$ galaxy environments due to their low metallicity (see e.g. \citealt{Hopkins2004AJ....128.1112H, Maiolino2004A&A...420..889M, Gallerani2010A&A...523A..85G}). To be able to fit the quasar spectra to the observed photometry we convolve the model spectra with the available photometric filter transmission profiles, which includes absorption by the atmosphere. Subsequently, the template scaling is calculated analytically across all bands. To determine the best fitting model a minimum $\chi^2$ is calculated for each template with $E(B-V)$ values in the range of $-$0.3-1.0 (with a step size of 0.005). We also fit negative values of $E(B-V)$ to allow for intrinsically bluer quasars than average. To compare the preference for each model we calculate the Bayesian Information Criterion (BIC), which is given by
\begin{equation}
    \text{BIC} = \chi^2 + k \ln N,
\end{equation}
in the case of Gaussian distributed errors, with $k$ the number of model parameters and $N$ the number of photometric data points. When comparing two models, a $\Delta$BIC > 6 or $\Delta$BIC < -6 gives a strong positive or negative evidence, respectively, for one of the models being preferred. Therefore, we compare the best fitting dust extinction models to the \cite{Selsing2016A&A...585A..87S} composite quasar spectrum fit and assign a $E(B-V)$ value of zero if this condition is not met. Furthermore, we do not include quasars with 3 or less photometric data points, to ensure a reliable fit.

The resulting $E(B-V)$ values using the SMC extinction law are shown in Fig.~\ref{fig:e_b_v_plot}. Only 1 quasar, ILTJ0121+2940, satisfies the criterion $E(B-V) > 0.1$ for both possible redshifts $z=5.24$ and $5.34$ as discussed in Section ~\ref{subsubsection:ILTJ0121+2940}. However, to examine the $E(B-V)$ values of our radio selected quasars compared to optically selected high-$z$ quasar sample, we repeat the same fitting procedure using the $z>5$ quasar sample of \cite{Ross2020MNRAS.494..789R}, which is also shown in Fig.~\ref{fig:e_b_v_plot}. We divide the literature sample in radio-loud (RL) and radio-quiet (RQ) using the record of known RL quasars (see \citealt{Banados2021ApJ...909...80B} and \citealt{Gloudemans2021A&A...656A.137G}). We note that the radio-quiet sample likely still contains a small fraction of RL quasars, which have not been identified yet. The fraction of RL quasars with $E(B-V) > 0.1$ (including quasars from this work) is 2.3\% for our quasars compared to 1.5\% for the RQ literature sample. Comparing the two distributions with a KS-test results in a p-value of 0.5, indicating no significant deviation. However, the sample size of RL quasars is still relatively small with 44 quasars and biased by observational selection. Including also the fits using the starburst galaxy extinction curve of \cite{Calzetti2000ApJ...533..682C} yields higher fractions of red quasars of 9.1\% and 3.2\% for the RL and RQ sample, respectively. However, there is still no significant deviation between the two distributions (p-value of 0.6), therefore supporting the same conclusion.

The fraction of possible red quasars we find in our sample is similar to \cite{Kato2020PASJ...72...84K}, who report a fraction of 2\% at $z>5.6$ using the Subaru High-$z$ Exploration of Low-Luminosity Quasars (SHELLQs; \citealt{Matsuoka2016ApJ...828...26M}) sample of 93 high-$z$ quasars. However, due to selection biases, observational limits, and definitions, red quasars are understudied at high redshift and likely biased towards lower fractions. 

At lower redshift \cite{Glikman2012ApJ...757...51G} have reported a red quasar fraction of $\sim20$\% between $0.13 < z < 3.1$ using 51 radio-selected quasars and \cite{Richards2003AJ....126.1131R} found a fraction of 15\% between $0.3 < z < 2.2$ using a sample of 4576 Sloan Digital Sky Survey (SDSS; \citealt{York2000AJ....120.1579Y}) quasars. Recently, \cite{Glikman2022ApJ...934..119G} have investigated the extinction properties of a large sample of SDSS QSOs at $0.1 < z \lesssim 3$ using a similar template fitting method as was used in this work. They find that the fraction of red quasars is strongly luminosity dependent, with red quasars making up to 40\% of the sources at the highest luminosities, but decreasing to a few percent at lower luminosities. To compare their findings to our sample, we plotted the $E(B-V)$ distribution found by \cite{Glikman2022ApJ...934..119G} at $z<3$ in Fig.~\ref{fig:e_b_v_plot}, which shows the distribution is similar in shape. However, we do not find the same tail of red quasars out to $E(B-V) \sim 1.0$ as \cite{Glikman2022ApJ...934..119G}, which is as expected since most red quasars would disappear from high-$z$ optical and NIR surveys due to the large amount of extinction. 

To investigate whether the Legacy Surveys depth allows for discovering heavily reddened quasars, we determine the maximum $E(B-V)$ for each quasar in our sample for which it would still be visible in the Legacy $z$-band (reaching a 5$\sigma$ depth of $m_z \sim 23.3$). The fitting procedure is similar to the one described above, but here we use only the $W1$ and $W2$ measurements of each quasar and increase the $E(B-V)$ value (using the SMC extinction law) on the best fitting quasar template until the Legacy $z$-band depth is reached. This analysis shows that 44\% of the quasars would not have been detected if they had $E(B-V)>0.25$ and 93\% if they had $E(B-V)>0.4$, which is consistent with not finding heavily reddened quasars in a tail out to $E(B-V) \sim 1.0$ and likely contributing to our lower fractions of red quasars. Therefore, we need significantly deeper optical data to detect the dust obscured quasars that are currently too faint for ground based telescopes. The Rubin Observatory Legacy Survey of Space and Time (LSST, \citealt{lsst2009arXiv0912.0201L}) will eventually reach a 5$\sigma$ depth of $m_z \sim 26.1$, which would yield a $>$80\% completeness out to $E(B-V)=0.4$. Furthermore, to study red quasar fractions as function of radio luminosity in a solid statistical way at high redshift we need a larger sample size of radio bright quasars, which can be discovered by combining future large area radio surveys from e.g. LOFAR and SKA with deep optical and NIR data from e.g. LSST and the forthcoming Euclid mission \citep{euclid2019A&A...631A..85E}.

\section{Summary and future prospects}
\label{sec:conclusions}

We have started a campaign to search for radio-bright high-$z$ quasars by combining an optical colour cut technique using the Legacy Surveys with LoTSS-DR2 radio detections. The LoTSS-DR2 survey is currently reaching sensitivities within the 100 $\mu$Jy regime, which enables detection of $\sim$30\% of the high-$z$ quasar population at low-frequencies \citep{Gloudemans2021A&A...656A.137G}. We have performed spectroscopic follow-ups for 80 of our candidates, which lead to the confirmation of 20 new quasars (and the independent confirmation of four recent confirmed high-$z$ quasars) between $4.9 \leq z \leq6.6$ and a doubling of the sample of known radio-loud quasars in this epoch. The radio detections from LoTSS-DR2 decreased the contamination of stellar sources in our sample and allowed for selection of these quasars in a broad redshift range and use of the limited number of optical bands ($g,\ r,$ and $z$ bands) from the Dark Energy Spectroscopic Instrument (DESI) Legacy Imaging Surveys, which probe at least 1 magnitude deeper than other large surveys such as PS1. Our campaign demonstrates the potential for efficiently discovering new (faint) high-$z$ quasar populations through next generation radio continuum surveys. 

Our new quasar sample probes on average fainter rest-frame UV luminosities and brighter radio luminosities than the previously known quasar population. The optical and NIR colours and Ly$\alpha$ EW distribution are very similar to the known quasar population, which potentially suggests that there is no strong radio-loudness dichotomy between the radio-loud and radio-quiet quasar population. However, we note that we find a higher fraction (38\%) of weak line quasars compared to the 14\% reported by \cite{Banados2016ApJS..227...11B}. Furthermore, we find little to no correlation between the Ly$\alpha$ and radio luminosity of our sample with a Pearson correlation coefficient of $r=-0.12$. However, this analysis is limited by the small sample size and observational selection biases.

Furthermore, we estimated the red quasar fraction of the RL versus the RQ high-$z$ quasar population by fitting the quasar composite of \cite{Selsing2016A&A...585A..87S} to the photometry and applying the SMC extinction law of \cite{Pei1992ApJ...395..130P} and starburst galaxy extinction curve of \cite{Calzetti2000ApJ...533..682C}. Our analysis suggests no significant deviation between the $E(B-V)$ values of the RL and RQ high-$z$ quasar sample.

We find that the success rate of our spectroscopic follow-up campaign can be improved significantly by decreasing the positional optical-radio separation limit of our candidates; by taking the optical-radio separation criterion as 1.2$\arcsec$ instead of 2.0$\arcsec$, the 34\% success rate could have been increased to 42\% without loss of new or known quasars. From a simple Monte-Carlo simulation we derive the probability of a quasar having an optical-radio separation larger than 1$\arcsec$ to be $\sim2$\%, assuming similar positional uncertainties. The separations should therefore be carefully considered in future work. However, note that the optimal separation limit is highly dependent on the resolution, sensitivity, calibration of the radio data, and target elevation (which causes changes in the synthesised beam). 

To further study the AGN and host galaxy properties of these new high-$z$ quasars, additional observations are necessary (see e.g. \citealt{Khusanova2022A&A...664A..39K}). The SMBH masses and broad line regions for example can be constrained by measuring the C\textsc{iv} and Mg\textsc{ii} NIR lines (see e.g. \citealt{Fine2010MNRAS.409..591F,Onoue2019ApJ...880...77O}) and host galaxy kinematics by measuring the sub-millimetre [C\textsc{ii}] emission line (e.g. \citealt{Neeleman2021ApJ...911..141N}). Furthermore, the radio spectra need to be constrained to be able to study the radio emission mechanisms of the AGN. A statistical sample of radio spectra of high-$z$ quasars is also essential to obtain better estimates of the number counts of radio sources suitable for future 21-cm experiments such as SKA \citep{ciardi2015aska.confE...6C} and to further optimise strategies for discovering suitable sources. 

A novel approach to finding high-$z$ radio sources will be undertaken by the WEAVE-LOFAR (WL) survey \citep{smith2016sf2a.conf..271S}. The WEAVE instrument \citep{Dalton2014SPIE.9147E..0LD} is a multi-fiber spectrograph on the William Herschel Telescope on La Palma and is expected to have first light in the summer of 2022. The WEAVE-LOFAR survey will obtain $\sim$1 million spectra of radio sources from the LoTSS surveys without optical colour cuts and will therefore probe the full range of potential quasar colours, as well as build large samples. This survey has the potential of enabling the detection of dust obscured high-$z$ radio galaxies and quasars into the EoR and discovery of suitable sources for 21-cm spectroscopy.

\begin{acknowledgements}
{For the purpose of open access, the author has applied a Creative Commons Attribution (CC BY) licence to any Author Accepted Manuscript version arising from this submission. We are grateful for the support of Kentaro Aoki and Ichi Tanaka during our Subaru/FOCAS observations (proposal ID S21B-003).
KJD acknowledges funding from the European Union's Horizon 2020 research and innovation programme under the Marie Sk\l{}odowska-Curie grant agreement No. 892117 (HIZRAD). 
Y.H. is supported by JSPS KAKENHI Grant Number 21K13953.

MJH and DJBS acknowledge support from the UK Science and Technology Facilities Council (STFC) under grant ST/V000624/1.

PNB is grateful for support from the UK STFC via grant ST/V000594/1.

The Low Resolution Spectrograph 2 (LRS2) was developed and funded by the University of Texas at Austin McDonald Observatory and Department of Astronomy, and by Pennsylvania State University. We thank the Leibniz-Institut fur Astrophysik Potsdam (AIP) and the Institut fur Astrophysik Goettingen (IAG) for their contributions to the construction of the integral field units.

We thank the Board of the Hobby-Eberly Telescope for the allocation of Guaranteed Time for the LRS2 instrument, which was important in enabling this investigation.

This paper is based (in part) on data obtained with the International LOFAR Telescope (ILT) under project codes LC0 015, LC2 024, LC2 038, LC3 008, LC4 008, LC4 034 and LT10 01. LOFAR \citep{vanHaarlem2013A&A...556A...2V} is the Low Frequency Array designed and constructed by ASTRON. It has observing, data processing, and data storage facilities in several countries, which are owned by various parties (each with their own funding sources), and which are collectively operated by the ILT foundation under a joint scientific policy. The ILT resources have benefited from the following recent major funding sources: CNRS-INSU, Observatoire de Paris and Universit\'e d'Orl\'eans, France; BMBF, MIWF-NRW, MPG, Germany; Science Foundation Ireland (SFI), Department of Business, Enterprise and Innovation (DBEI), Ireland; NWO, The Netherlands; The Science and Technology Facilities Council, UK; Ministry of Science and Higher Education, Poland.

This research made use of the Dutch national e-infrastructure with support of the SURF Cooperative (e-infra 180169) and the LOFAR e-infra group. The J\"ulich LOFAR Long Term Archive and the German LOFAR network are both coordinated and operated by the J\"ulich Supercomputing Centre (JSC), and computing resources on the supercomputer JUWELS at JSC were provided by the Gauss Centre for Supercomputing e.V. (grant CHTB00) through the John von Neumann Institute for Computing (NIC).

This research made use of the University of Hertfordshire high-performance computing facility and the LOFAR-UK computing facility located at the University of Hertfordshire and supported by STFC [ST/P000096/1], and of the Italian LOFAR IT computing infrastructure supported and operated by INAF, and by the Physics Department of Turin university (under an agreement with Consorzio Interuniversitario per la Fisica Spaziale) at the C3S Supercomputing Centre, Italy.

}
	
\end{acknowledgements}

\bibliographystyle{aa}
\bibliography{bibliography.bib}

\begin{appendix}

\section{Contaminants}
\label{sec:appendix_contaminants}
All spectroscopically observed candidates that have been identified as not being quasars are listed in Table~\ref{tab:contaminants} for future reference. Based on initial visual inspection, the majority ($\sim$60\%) of the contaminants we have detected in our spectroscopic follow-up observations are M-type brown dwarf (BD) candidates. If these are indeed proven to be BDs with associated radio emission, this is an interesting discovery by itself, since the first direct low-frequency radio detections of BDs have only recently been reported by \cite{Vedantham2020ApJ...903L..33V} and \cite{Callingham2021NatAs...5.1233C}. In those studies, their radio emission can partially be explained by plasma emission from the active chromospheres of the stars. However, the origin of radio emission for the in-active stars is potentially driven by star-planet interactions. The investigation of the BD nature of these sources is beyond the scope of this paper and will be discussed in separate work (Gloudemans et al. in prep.).

The other $\sim$40\% of the contaminants have slowly rising featureless spectra and are expected to be either low-$z$ SF galaxies with possible AGN or L- or T-type dwarfs, which are difficult to classify in the optical wavelength regime. In the case of low-$z$ SF galaxies the red-dusty spectrum could be confused with a high-$z$ quasar SED due to a strong Balmer break (around $z\sim$1-2) or dust extincted continuum (see e.g. \citealt{McLure2010MNRAS.403..960M, Donley2012ApJ...748..142D, Dunlop2013ASSL..396..223D}). L- and T-type dwarfs also exhibit a slowly rising spectrum in the optical, however to classify these follow-up IR observations are needed to identify for example the TiO and VO lines and methane absorption (see e.g. \citealt{Knapp2004AJ....127.3553K}). 

\begin{table}[h]
\caption{Spectroscopically observed candidates that are confirmed not to be high-$z$ quasars.} 
\label{tab:contaminants}
\centering
\resizebox{\columnwidth}{!}{
\begin{tabular}{c c c c c }  
\hline\hline       
Name & RA (hms) & DEC (dms) & Radio-optical & Telescope/ \\ 
 & (J2000) & (J2000) & sep. (arcsec) & instrument \\ 
\hline 

ILTJ0010+3239 & 00:10:08.75 & +32:39:01.18 & 0.7 & Subaru \\
ILTJ0012+3403 & 00:12:11.02 & +34:03:50.32 & 1.5 & HET \\
ILTJ0056+3200 & 00:56:45.16 & +32:00:48.51 & 0.4 & Subaru \\
ILTJ0116+2535 & 01:16:48.39 & +25:35:01.18 & 2.0 & VLT \\
ILTJ0220+2520 & 02:20:46.06 & +25:20:26.68 & 0.9 & Subaru \\
ILTJ0729+3809 & 07:29:03.37 & +38:09:45.24 & 1.1 & Subaru \\
ILTJ0733+3159 & 07:33:59.10 & +31:59:46.26 & 0.5 & Subaru \\
ILTJ0754+3456 & 07:54:22.59 & +34:56:26.14 & 0.3 & HET \\
ILTJ0754+3610 & 07:54:35.60 & +36:10:31.17 & 1.9 & HET \\
ILTJ0802+2523 & 08:02:12.50 & +25:23:56.00 & 0.5 & Subaru \\
ILTJ0821+4831 & 08:21:47.96 & +48:31:00.13 & 1.7 & HET \\
ILTJ0828+4203 & 08:28:04.87 & +42:03:46.75 & 0.8 & HET \\
ILTJ0831+2633 & 08:31:29.01 & +26:33:34.46 & 1.0 & VLT \\
ILTJ0909+4502 & 09:09:31.06 & +45:02:01.81 & 1.7 & HET \\
ILTJ0933+5215 & 09:33:42.71 & +52:15:17.83 & 1.0 & HET \\
ILTJ0933+5248 & 09:33:59.67 & +52:48:02.84 & 1.7 & HET \\
ILTJ0950+3440 & 09:50:21.02 & +34:40:29.06 & 1.0 & HET \\
ILTJ1004+3424 & 10:04:10.77 & +34:24:34.12 & 0.7 & Subaru \\
ILTJ1010+4234 & 10:10:54.32 & +42:34:39.90 & 1.4 & Subaru \\
ILTJ1011+5757 & 10:11:55.86 & +57:57:36.55 & 0.4 & Subaru \\
ILTJ1012+6532 & 10:12:17.72 & +65:32:10.90 & 1.4 & Subaru \\
ILTJ1020+5909 & 10:20:01.89 & +59:09:01.96 & 0.9 & HET \\
ILTJ1039+4936 & 10:39:35.52 & +49:36:24.22 & 0.7 & Subaru \\
ILTJ1056+3112 & 10:56:15.94 & +31:12:32.83 & 0.6 & Subaru \\
ILTJ1056+4759 & 10:56:23.66 & +47:59:06.56 & 0.7 & Subaru \\
ILTJ1059+5516 & 10:59:22.43 & +55:16:21.10 & 0.7 & Subaru \\
ILTJ1101+6507 & 11:01:31.65 & +65:07:50.99 & 0.8 & Subaru \\
ILTJ1124+5141 & 11:24:58.70 & +51:41:16.79 & 0.6 & HET \\
ILTJ1132+3651 & 11:32:13.29 & +36:51:49.93 & 0.1 & HET \\
ILTJ1135+3336 & 11:35:04.90 & +33:36:39.98 & 1.2 & HET \\
ILTJ1354+4234 & 13:54:51.40 & +42:34:01.85 & 1.3 & HET \\
ILTJ1357+3354 & 13:57:30.24 & +33:54:36.17 & 0.7 & HET \\
ILTJ1402+2914 & 14:02:02.45 & +29:14:14.17 & 1.1 & HET \\
ILTJ1519+4750 & 15:19:09.56 & +47:50:26.97 & 1.5 & HET \\
ILTJ1535+6145 & 15:35:46.82 & +61:45:59.39 & 0.9 & Keck \\
ILTJ1755+3937 & 17:55:45.66 & +39:37:44.23 & 0.5 & HET \\
ILTJ1801+3141 & 18:01:08.08 & +31:41:13.48 & 2.0 & HET \\
ILTJ2157+2108 & 21:57:22.41 & +21:08:12.90 & 0.3 & VLT \\
ILTJ2210+2449 & 22:10:21.05 & +24:49:52.49 & 0.7 & Subaru \\
ILTJ2232+1840 & 22:32:16.06 & +18:40:04.28 & 0.6 & HET \\
ILTJ2235+1802 & 22:35:25.17 & +18:02:37.67 & 1.2 & Subaru \\
ILTJ2244+2140 & 22:44:48.02 & +21:40:54.13 & 1.7 & Subaru \\
ILTJ2259+2018 & 22:59:57.36 & +20:18:49.44 & 0.9 & HET \\
ILTJ2308+1925 & 23:08:30.46 & +19:25:59.74 & 0.4 & Subaru \\
ILTJ2316+3142 & 23:16:09.78 & +31:42:40.18 & 1.4 & Subaru \\
ILTJ2327+2002 & 23:27:56.67 & +20:02:50.45 & 1.9 & HET \\
\hline 
\vspace{0.5pt}
\end{tabular}}
\end{table}

\clearpage

\section{Cutouts of newly discovered quasars}
\label{sec:appendix_cutouts}

\begin{figure}[p]
\vspace*{1cm}
\hspace*{-5cm}
\makebox[\linewidth]{
        \includegraphics[width=0.65\textwidth]{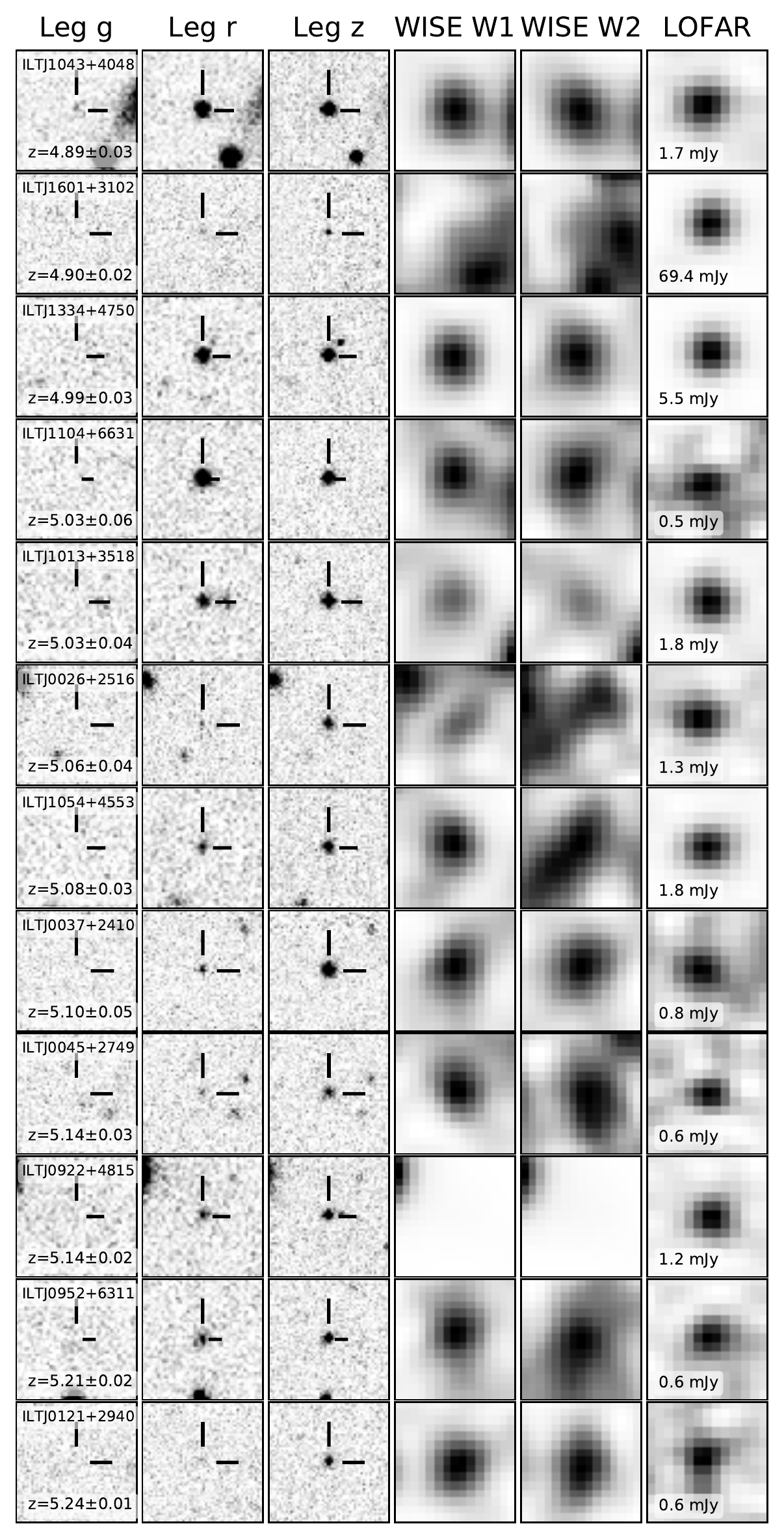}
    }
     \label{fig:cutouts}
\end{figure}

\begin{figure*}
\centering
   \includegraphics[width=0.65\textwidth, trim={0.0cm 0cm 0cm 0.0cm}, clip]{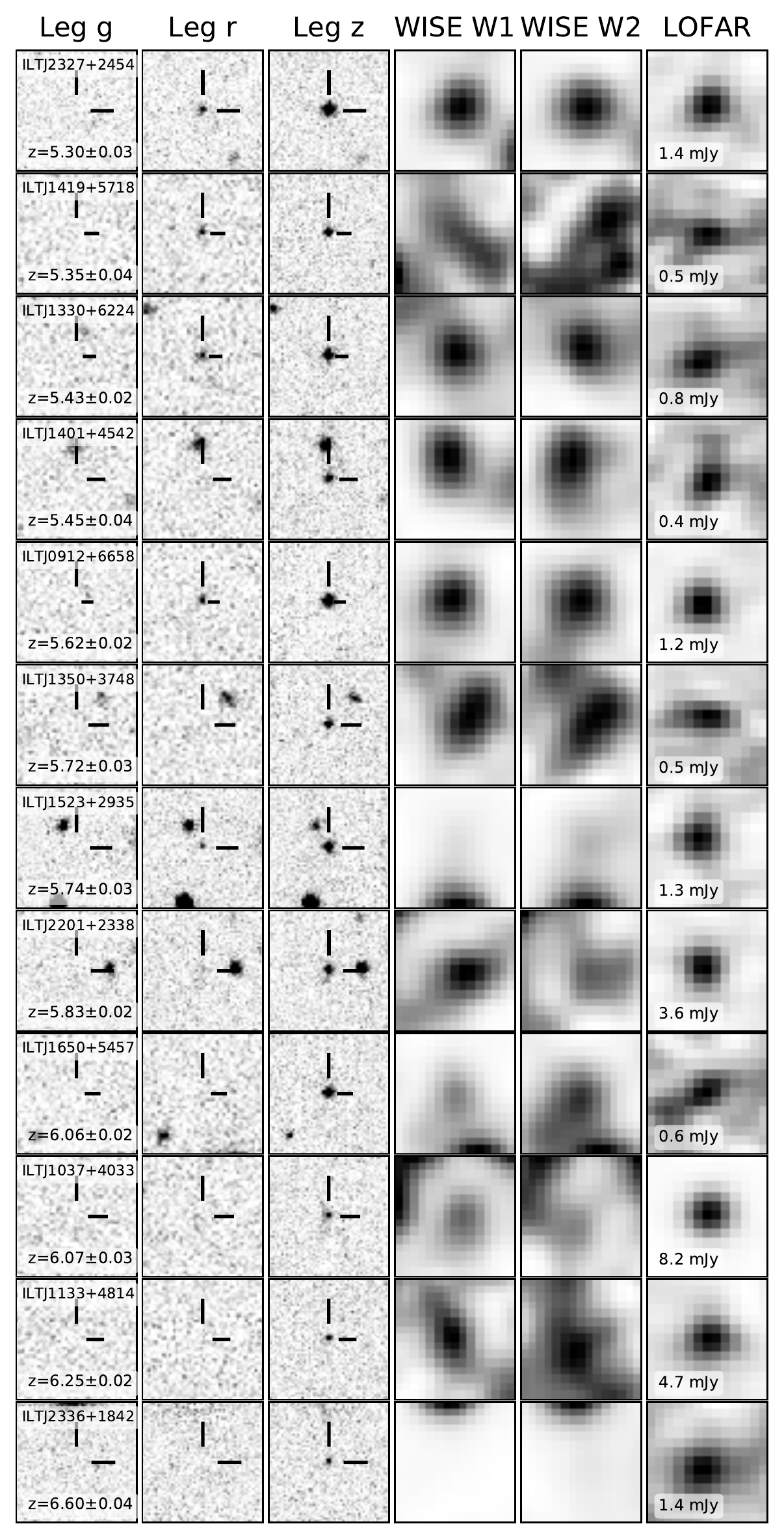}
     \caption{20"$\times$20" Legacy DR8 \citep{dey2019AJ....157..168D}, AllWISE \citep{Cutri2014yCat.2328....0C}, and LOFAR (LoTSS-DR2; \citealt{Shimwell2022A&A...659A...1S}) cutouts for the newly confirmed quasars.}
     \label{fig:cutouts2}
\end{figure*}

\end{appendix}

\end{document}